\documentclass[%
reprint,
superscriptaddress,
nofootinbib,
 amsmath,amssymb,
prb,
floatfix
]{revtex4-2}

\usepackage[dvipsnames]{xcolor}
\usepackage{graphicx}
\usepackage{dcolumn}
\usepackage{bm}
\usepackage[colorlinks=true,linkcolor=Blue,citecolor=Blue,urlcolor=Blue]{hyperref}
\usepackage[T1]{fontenc}
\usepackage{subdepth}
\usepackage{physics}
\usepackage{bbm}
\usepackage{setspace}
\usepackage{mathtools}
\usepackage{array}

\usepackage{cleveref}

\crefname{equation}{}{}
\Crefname{equation}{}{}
\crefrangelabelformat{equation}{(#3#1#4--#5#2#6)}
\newlength{\arrow}
\allowdisplaybreaks
\begin{document}

\title{Coherence limitations in the optical control of the singlet-triplet qubit in a quantum dot molecule}
\date{\today}
\author{Karol Kawa}
\email{Karol.Kawa@pwr.edu.pl}
\affiliation{Department of Theoretical Physics, Wroc{\l}aw University of Science and Technology, 50-370, Wroc{\l}aw, Poland}
\affiliation{Institut f\"ur Festk\"orpertheorie, Westf\"alische Wilhelms-Universit\"at M\"unster, D-48149, M\"unster, Germany}
\author{Tilmann Kuhn}
\affiliation{Institut f\"ur Festk\"orpertheorie, Westf\"alische Wilhelms-Universit\"at M\"unster, D-48149, M\"unster, Germany}

\author{Pawe{\l} Machnikowski}
\affiliation{Department of Theoretical Physics, Wroc{\l}aw University of Science and Technology, 50-370, Wroc{\l}aw, Poland}

\begin{abstract}
We analyze the optically driven dynamics of a qubit implemented on a singlet-triplet subspace of two-electron  states in a self-assembled quantum dot molecule.
We study two possible control schemes based on the coupling to an excited (four-particle) state either by two spectrally separated laser pulses or by a single spectrally broad pulse.
We quantitatively characterize the imperfections of the qubit operation resulting from non-adiabatic evolution and from limited spectral selectivity in a real system, as compared to the ideal adiabatic Raman transfer of occupation in the $\Lambda$-system.
Next, we study the effects of decoherence induced by the coupling to the phonons of the surrounding crystal lattice and by radiative recombination.
As a result, we are able to identify the optimization trade-offs between different sources of errors and indicate the most favorable conditions for quantum control of the singlet-triplet qubit in the two optical control schemes.
\end{abstract}

\maketitle

\section{Introduction}

One of the challenges of quantum computation and networking is to build a quantum interface between the computational registers and quantum communication links. Solid-state systems, like carrier spins in self-assembled semiconductor quantum dots (QDs), offer a viable way toward this goal \cite{Yilmaz2010,DeGreve2012,Gao2012} by providing relatively stable quantum registers with lifetimes reaching seconds \cite{Kroutvar2004,Gillard2021} and coherence times of the order of $\mu$s \cite{Stockill2016}, combined with a high level of optical control, including high-fidelity single spin initialization \cite{Atature2006,Xu2007,Kim2008}, fast spin manipulation \cite{Berezovsky2008, Ramsay2008, Press2008}, and non-destructive readout \cite{Kim2008,Atatre2007}. 
An implementation of the quantum bit is also possible in an artificial molecule composed of two coupled QDs (quantum dot molecule, QDM) with the benefit of the reduced impact of fluctuating magnetic environment and charge fluctuations, hence extended coherence times \cite{Weiss_PRL12}.
Here, the qubit space is spanned by the singlet and triplet states of two electrons with a vanishing $z$-component of the total spin.
These states are both optically coupled to the same four-particle configuration, in which an additional electron-hole pair (exciton) is created in one of the QDs \cite{Greilich2011,Kim2011,Weiss_PRL12}.
This renders optical control of such a singlet-triplet qubit possible in the standard frame of a $\Lambda$-system.
Entanglement with photons makes it possible to couple the qubit to quantum communication lines \cite{Delley2017,Awschalom2018}, while resonance fluorescence techniques offer potential toward single-shot readout \cite{Delley2015a,Farfurnik2021}.



While spin states are relatively stable, the absence of a substantial coupling between light and spin forces the optical spin control schemes to rely on spin-dependent charge dynamics involving the coupling to an excited state \cite{Pazy2003,chen04,Grodecka2007,Kim2011}.
This makes the qubit vulnerable to errors that may be due to the occupation leakage to the ``virtually'' coupled auxiliary state, its decay due to radiative recombination, as well as to the dynamical response of the crystal lattice, i.e., phonons, which leads to dephasing and affects the fidelity of the qubit operation \cite{Roszak2005,Caillet2007,Grodecka2007}.

In this paper, we study two possible ways of optical rotation of a singlet-triplet qubit and analyze the leakage and dephasing channels that limit the fidelity of the quantum gating protocol. The first control method to be discussed follows the standard scheme of inducing an arbitrary spin rotation via an adiabatic Raman transfer with two simultaneous, spectrally selective laser pulses off-resonantly coupled to the excited state \cite{chen04,Grodecka2007}.
The second one, experimentally implemented in Ref.~\cite{Kim2011}, uses a single spectrally broad pulse to couple both triplet and singlet states to the excited state.
We analyze the imperfections of the evolution with respect to the intended adiabatic control scheme and discuss the effects of environmentally induced decoherence as a function of the system and control parameters.
We show that simultaneous optimization of the fidelity against all sources of error is possible within the parameter space for the two-color, two-pulse scheme, while in the single-color, single-pulse protocol a trade-off between different error mechanisms has to be resolved. 

The organization of the paper is as follows.
First, in Sec.~\ref{sec:model-system} we introduce the physical system under study and the model describing it.
Next, in Sec.~\ref{sec:gating-protocol} we present the two protocols for qubit rotation.
Sec.~\ref{sec:leakage-of-quantum-information} focuses on the  unwanted effect of leakage of quantum information into the auxiliary state and on the imperfections of the performed quantum gate.
Next, we focus on environmentally induced decoherence processes. 
In Sec.~\ref{sec:general-theory of-error}, we summarize a general theory of the decoherence mechanism based on the perturbation theory of the density matrix.
In Sec.~\ref{sec:phonon-bath} we employ this theory  to the impact of phonons on the fidelity of the quantum gate, while in Sec. \ref{sec:photon-bath} the errors induced by radiative recombination are analyzed.
Finally, in Sec.~\ref{sec:summary-and-discussion}, we conclude our work and discuss the possibility of minimizing the error within the available space of control parameters.

\section{The system and the model \label{sec:model-system}}

\begin{figure}[t]
    \centering
    \includegraphics[width=0.8\linewidth]{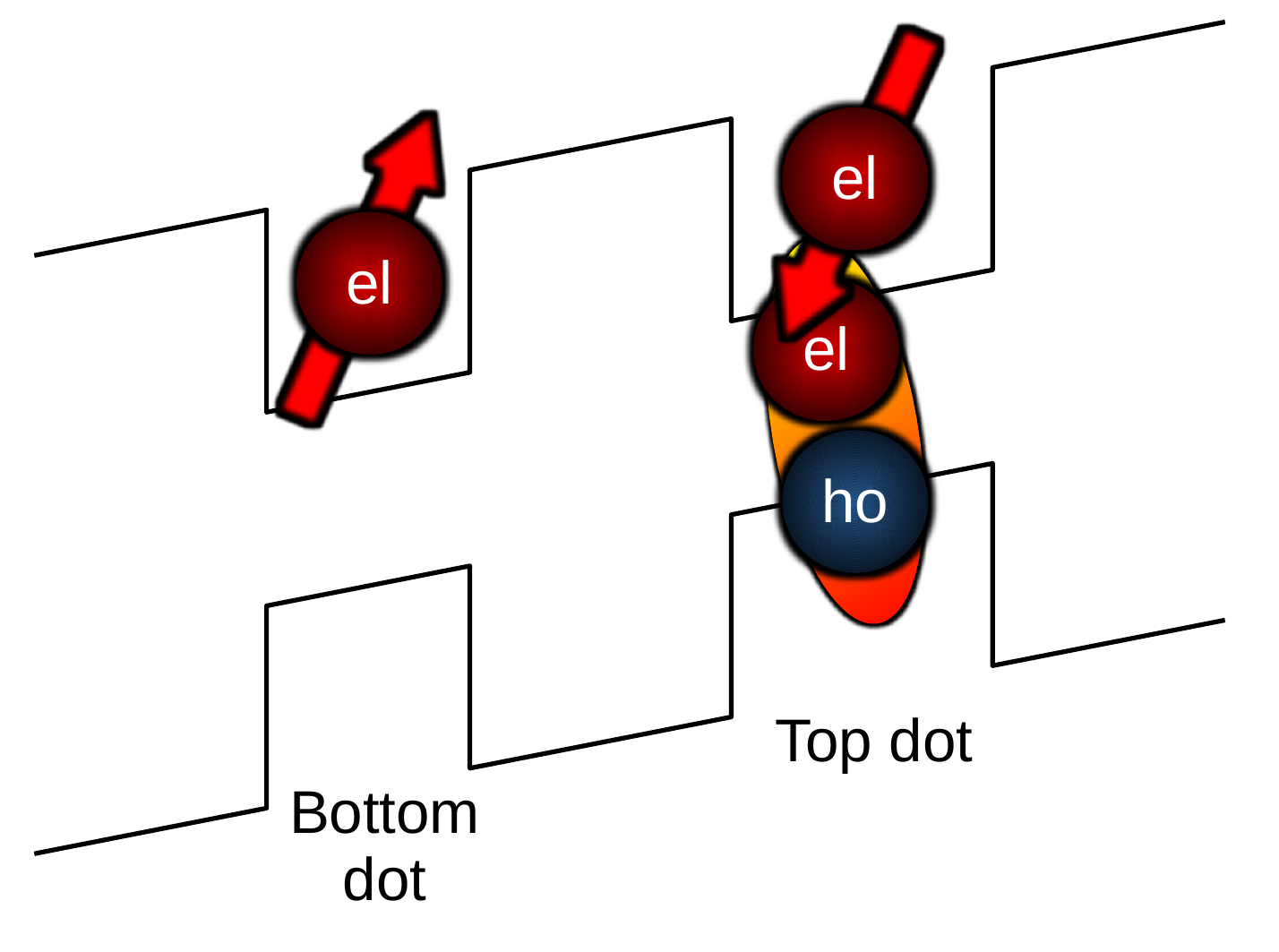}
    \caption{Illustrative representation of the system in the form of the band edge diagram: a QDM is occupied by two electrons. Optical excitation couples the two-electron states to an excited state with an addition exciton created in the upper dot. Here the direction of the epitaxial growth of the structure (the ``vertical'' $z$ axis) is from left to right; the electric field is applied in this direction.} 
    \label{fig:system}
\end{figure}

We consider a QDM formed by two vertically stacked self-assembled InAs QDs, where, to be specific, we assume that the upper QD is larger resulting in a smaller excitation energy than the lower one.
The QDM is placed in a field effect (diode) structure \cite{Greilich2011,Kim2011} to provide controlled charging (Fig.~\ref{fig:system}).
We assume that the QDM is charged by two resident electrons.
In a minimal model, which is sufficient for our purpose, the electrons can occupy the lowest states in each QD.
We assume that the external bias voltage is set to such a value that the lowest-energy two-electron states correspond to singly occupied QDs, while the doubly occupied configurations are energetically higher due to Coulomb blockade.
The qubit subspace is then spanned by the singlet state
\begin{gather*}
    \ket*{S} = \dfrac{1}{\sqrt{2}}\left(a_{1\uparrow}^\dagger a_{2\downarrow}^\dagger - a_{1\downarrow}^\dagger a_{2\uparrow}^\dagger\right)\ket{\mathrm{vac}},
\end{gather*}
and the triplet state with zero projection of the angular momentum, i.e., $J_z=0$,
\begin{gather*}
    \ket*{T} = \dfrac{1}{\sqrt{2}}\left(a_{1\uparrow}^\dagger a_{2\downarrow}^\dagger + a_{1\downarrow}^\dagger a_{2\uparrow}^\dagger\right)\ket{\mathrm{vac}}.
\end{gather*}
Here $\ket*{\mathrm{vac}}$ denotes the state of empty molecule and $a_{j,s}^\dagger$ is the electron creation operator in the $j$-th QD ($j=1,2$) with the spin $s=\uparrow,\downarrow$.
The states $\ket*{T}$ and $\ket*{S}$ are split by the exchange coupling which is tunable within a certain range of the electric field.
A comprehensive study of two-electron states along with the excited states in an artificial molecule can be found, e.g., in Ref.~\cite{Doty2008}.

\begin{figure}[t]
    \centering
    \begin{minipage}{0.49\linewidth}
        \includegraphics[width=\linewidth]{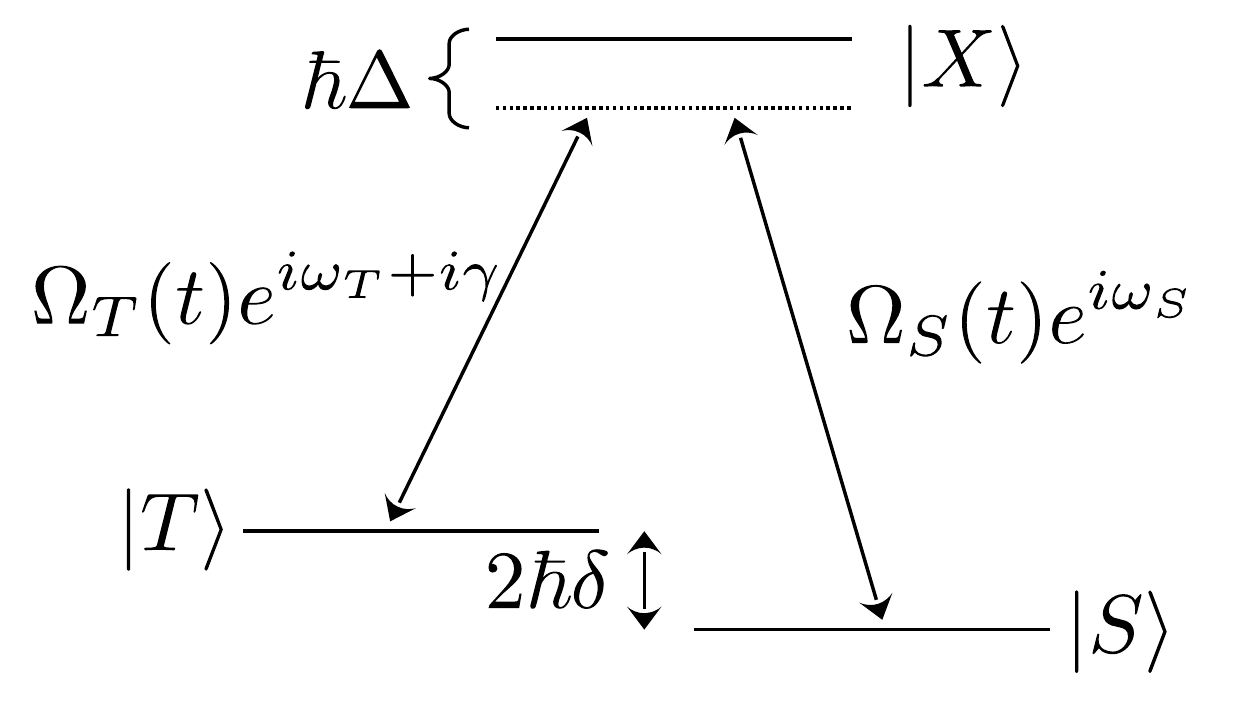}
        \small{(a)}
    \end{minipage}
    \begin{minipage}{0.49\linewidth}
        \includegraphics[width=\linewidth]{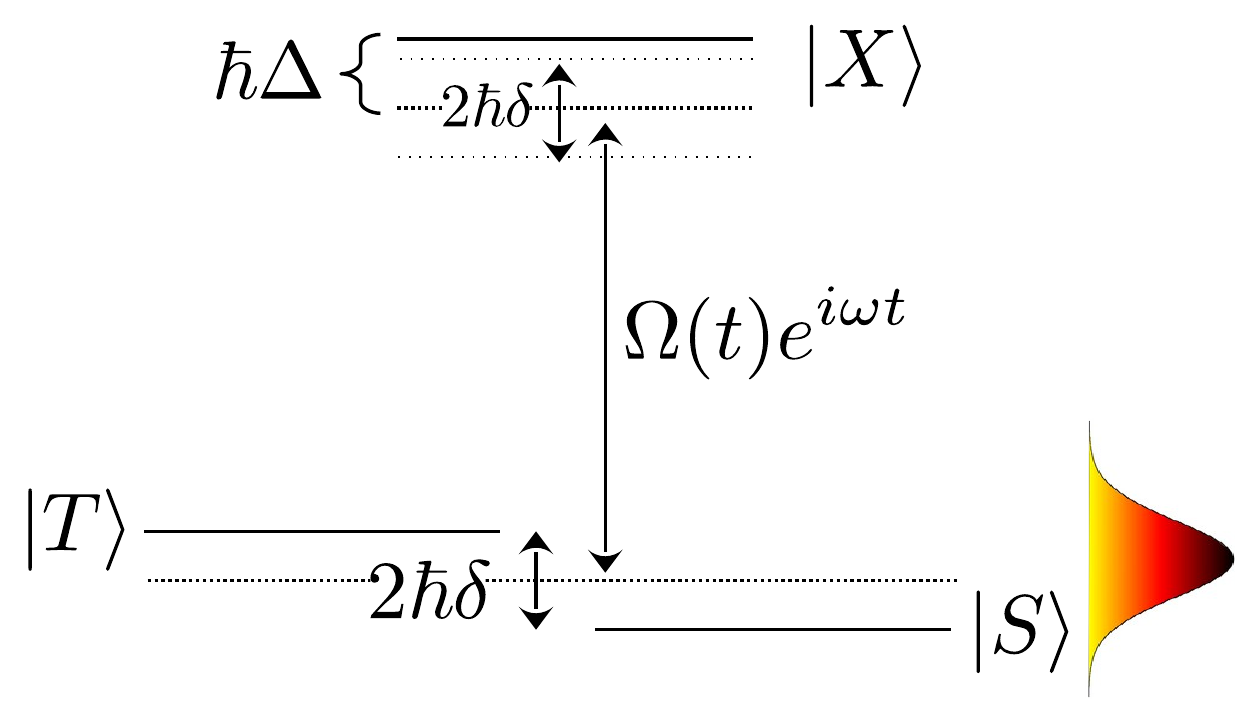}
        \small{(b)}
    \end{minipage}
    \caption{(a) The $\Lambda$-system necessary for optical control of singlet-triplet qubit using two laser pulses selectively coupling singlet and triplet to the exciton state with Raman conditions satisfied. (b) The $\Lambda$-system for optical spin control using only one broad laser pulse coupling both triplet and singlet to $\ket*{X}$.}
     \label{fig:lambda-system}
\end{figure}

Both qubit states are coupled to a four-particle configuration with angular momentum $J_z=-1$ via an optical transition induced by a $\sigma_{-}$ circularly polarized light field that creates an additional electron-hole pair in the upper QD \cite{Kim2011}
\begin{gather*}
    \ket*{X} = h_{2\Downarrow}^\dagger a_{2\uparrow}^\dagger a_{1\uparrow}^\dagger a_{2\downarrow}^\dagger \ket*{\mathrm{vac}}.
\end{gather*}
Due to Coulomb binding of the electron-hole pair and Coulomb blockade against double charging of the QDs, this configuration is stable in a certain range of electric fields.  
The three states $\ket*{T}$, $\ket*{S}$ and $\ket*{X}$ form a three-level $\Lambda$-system (see Fig.~\ref{fig:lambda-system}), which allows one to control the  $\ket*{T}$ and $\ket*{S}$ states by a detuned optical coupling to the $\ket*{X}$ state.
It should be noted that the triplet states with angular momentum projections $J_z = \pm 1$ are only coupled to four-particle states with $J_z = \pm 2$ and therefore do not interfere with the present $\Lambda$-system. 
Furthermore, by using the Zeeman effect, they can be shifted to different energies.

The system is described by the Hamiltonian
\begin{gather*}
    H = H_\mathrm{c} + H_{\mathrm{env}} + V_{\textrm{c-env}},
\end{gather*}
where the three components account for the confined carriers coupled to a laser field, the environment and the interaction between them.

The carrier-laser Hamiltonian describes the three-level $\Lambda$-system formed by the two-electron triplet and singlet states coupled to the lowest 4-particle state $\ket*{X}$ via a classical light beam, as presented in Fig.~\ref{fig:lambda-system}. It has the form
\begin{equation*}
    H_\mathrm{c} = \sum_{n=S,T,X} \epsilon_n \dyad{n}{n} + H_{L}.
\end{equation*}
Here $\epsilon_n$ are the energies of the two-particle states, with the exchange splitting $\epsilon_T-\epsilon_S=2\hbar\delta$, and the last term describes the carrier-light coupling for a $\sigma_-$ circular polarization of the laser beam in the rotating wave approximation,
\begin{equation}
    H_L = \dfrac{\hbar}{2} \bm{d}\cdot\bm{\mathcal{E}}^{(-)}(t) a_{2\uparrow}h_{2\Downarrow} + \mathrm{h.c.},
    \label{HL}
\end{equation}
where $\mathcal{E}^{(-)}(t)$ is the negative frequency part of the driving field. 

The environment Hamiltonian consists of two contributions,
\begin{gather*}
H_\mathrm{env} = H_\mathrm{ph} + H_\mathrm{rad},
\end{gather*}
corresponding, respectively, to the lattice (phonon) and radiative (photon) bath.
The free phonon Hamiltonian has the form
\begin{gather*}
    H_\mathrm{ph} = \sum_{\mathbf{k}} \hbar\omega_{\mathbf{k}} b_{\mathbf{k}}^\dagger b_{\mathbf{k}},
\end{gather*}
where $b^\dagger_\mathbf{k}$ ($b_\mathbf{k}$) is the bosonic creation (annihilation) operator for a phonon with wave vector $\mathbf{k}$ while $\omega_\mathbf{k}$ is the phonon frequency, which can be expressed by the wave vector and the velocity of phonons $c_l$ by $\omega_\mathbf{k} = c_l|\mathbf{k}|$. We will restrict the discussion to the deformation-potential coupling to longitudinal acoustic phonons, hence only this single phonon branch is relevant. 
The Hamiltonian of the radiative bath is
\begin{gather*}
    H_\mathrm{rad} = \sum_{\mathbf{q},\lambda} \hbar \omega_{\mathbf{q}}' c_{\mathbf{q}\lambda}^\dagger c_{\mathbf{q}\lambda}.
\end{gather*}
where $c^\dagger_{\mathbf{q}\lambda}$ ($c_{\mathbf{q}\lambda}$) is the bosonic creation (annihilation) operator of a photon with wave vector $\mathbf{q}$ and polarization index $\lambda$ and $\omega'_\mathbf{q} = c|\mathbf{q}|/n_\mathrm{r}$ is the photon frequency with $c$ as speed of light in vacuum and $n_\mathrm{r}$ as the refractive index of the crystal.

Finally, the interaction part of Hamiltonian can be written as
\begin{gather*}
    V = H_{\textrm{c-ph}} + H_{\textrm{c-rad}},
\end{gather*}
where $H_{\textrm{c-ph}}$ accounts for the interaction with phonons and $H_{\textrm{c-rad}}$ for the interaction with photons.
The phonon coupling Hamiltonian is of the form
\begin{equation*}
\begin{split}
&H_\textrm{c-ph} = \\
&=\left[\sum_{\mathbf{k}} f_2^{(\mathrm{h})}(\mathbf{k}) h_{2\Downarrow}^\dagger h_{2\Downarrow} + \sum_\mathbf{k} \sum_{j\sigma} f_j^{(\mathrm{e})} a_{j\sigma}^\dagger a_{j\sigma}\right] \left(b_\mathbf{k} + b_{-\mathbf{k}}^\dagger \right)
\end{split}
\label{eq:carrier-phonon-interactions}
\end{equation*}
where $f_j^{(\mathrm{e},\mathrm{h})}(\mathbf{k})$ is the coupling constant for the $j$th QD for electrons ($\mathrm{e}$) and holes ($\mathrm{h}$), respectively.
The form of this Hamiltonian reflects the facts that only one hole state is relevant, phonons cannot induce interband transitions due to huge energy mismatch, the coupling between the states localized in different QDs is inefficient due to small wave function overlap and spin-nonconserving phonon couplings are neglected because they correspond to weak  spin-orbit couplings.
Projected on the relevant subspace $\{\ket{T},\ket{S},\ket{X}\}$, this Hamiltonian takes the form
\begin{equation}
\begin{split}
    H_\textrm{c-ph} =& \sum_n \ev{H}{n} \dyad{n} \\
    =&\left\{  \sum_{\mathbf{k}} \left(\dyad{S} + \dyad{T} \right) \left[f_1^{(\mathrm{e})}\left(\mathbf{k}\right)+f_2^{(\mathrm{e})}\left(\mathbf{k}\right)\right]\right.\\
    &+ \left.\sum_\mathbf{k} \dyad{X} \left[f_1^{(\mathrm{e})}\left(\mathbf{k}\right)+2f_2^{(\mathrm{e})}\left(\mathbf{k}\right) + f_2^{(\mathrm{h})}\left(\mathbf{k}\right)\right] \right\} \\
    &\times \left(b_\mathbf{k} + b_{-\mathbf{k}}^\dagger\right).
\end{split}
\label{eq:hamiltonian-environment-phonons}
\end{equation}
The off-diagonal elements $\mel{S}{H}{T}$ vanish.
We carry out a canonical transformation to adjust the lattice equilibrium of the two resident electrons according to
\begin{gather*}
    b_\mathbf{k} = \tilde{b}_\mathbf{k} - \frac{f_1^{(\mathrm{e})}\left(\mathbf{k}\right) + f_2^{(\mathrm{e})}\left(\mathbf{k}\right)}{\hbar \omega_\mathbf{k}}.
\end{gather*}
Using the completeness relation $\dyad{T}+\dyad{S}+\dyad{X} = \mathbbm{1}$ we obtain
\begin{align}
\begin{aligned}
 H_\mathrm{ph} + &H_\textrm{c-ph} =\\ &=\sum_\mathbf{k} \left[\hbar \omega_\mathbf{k} \tilde{b}_\mathbf{k}^\dagger \tilde{b}_\mathbf{k} + \dyad{X}  F(\mathbf{k}) \left(\tilde{b}_\mathbf{k} + \tilde{b}_{-\mathbf{k}}^\dagger\right)\right],   
\end{aligned}
\label{eq:phonon-carrier-interaction-simpified}
\end{align}
where
\begin{equation*}
    F(\mathbf{k}) = f_2^{(\mathrm{e})}(\mathbf{k}) + f_2^{(\mathrm{h})}(\mathbf{k}).
\end{equation*}

Thus, only the excited state couples to phonons, which reflects the fact that the lattice responds to the change in the charge state of the system. The transformation leads also to a small shift of the energy $\epsilon_X$, the polaron shift, which in the following we take to be included in $\epsilon_X$.
The coupling constants have the form 
\begin{equation*}
    f_2^{(\mathrm{x})}\left(\mathbf{k}\right) = \sqrt{\frac{\hbar k}{2\rho \mathcal{V} c_l}}D_\mathrm{x} \mathcal{F}_\mathrm{x}\left(\mathbf{k}\right), \quad \mathrm{x} = \mathrm{e},\mathrm{h},
\end{equation*}
where $D_{\mathrm{e(h)}}$ is the deformation potential constant, $\mathcal{V}$ is the normalization volume, and 
$\mathcal{F}_\mathrm{x}\left(\mathbf{k}\right)$ is the form factor
\begin{gather*}
    \mathcal{F}_\mathrm{x}\left(\mathbf{k}\right) = \int\limits_{-\infty}^\infty \dd  \mathbf{r} \Psi_\mathrm{x}^*(\mathbf{r}) e^{i\mathbf{k}\cdot\mathbf{r}}\Psi_\mathrm{x}(\mathbf{r}).
\end{gather*}
For simplicity, we assume the same Gaussian form of the wave functions for the electron and the hole in the top quantum dot, i.e.,
\begin{gather*}
    \Psi_{\mathrm{e}}(\mathbf{r}) = \Psi_{\mathrm{h}}(\mathbf{r})
    \propto \exp\left(-\frac{r_\bot^2}{2l_\bot^2} - \frac{z^2}{2l_z^2}\right),
\end{gather*}
with $l_\bot$ and $l_z$ denoting the horizontal and vertical wave function widths, respectively.
With this choice, the form factor has also the same form for the electron and hole,
\begin{equation*}
    \mathcal{F}_{\mathrm{e}} \left(\mathbf{k}\right) = \mathcal{F}_{\mathrm{h}}\left(\mathbf{k}\right)
    = e^{-\left(\frac{k_\bot l}{2}\right)^2 - \left(\frac{k_z l_z}{2}\right)^2}.
\end{equation*}

The electron-photon interaction Hamiltonian is
\begin{gather}
    H_\textrm{c-rad} = \sum_{\mathbf{q}, \lambda} g_{\mathbf{q}\lambda} \left(c_{\mathbf{q}\lambda}+c_{\mathbf{-q}\lambda}^\dagger \right)
    \Bigl(\dyad{T}{X} + \dyad{S}{X} + \mathrm{h.c.}\Bigl )
    \label{eq:carrier-photon-interaction}
\end{gather}
with
\begin{gather*}
    g_{\mathbf{q}\lambda} = g^*_{-\mathbf{q}\lambda} = \sqrt{\frac{\hbar\omega'_\mathbf{q}}{2\epsilon_0\epsilon_\mathrm{r} \mathcal{V}}} \bm{d}\cdot  \bm{\hat{e}}^{(\lambda)} (\mathbf{q}),
\end{gather*}
where $\bm{d}$ is the interband dipole moment, $\bm{\hat{e}}^{(\lambda)} (\mathbf{q}) = \bm{\hat{e}}^{(\lambda)*} (-\mathbf{q})$ is the polarization unit vector, $\epsilon_0$ and $\epsilon_\mathrm{r}$ are the dielectric constants for vacuum and material semiconductor, respectively.

\section{The gating protocols\label{sec:gating-protocol}}

In this section, we present the spin rotation procedure, define the ideal (adiabatic) evolution of the system, and define the Hamiltonians that lead to non-adiabatic corrections. 
We first focus on the description of the two-color protocol in Sec.~\ref{sec:first-pro} and then summarize the single-color protocol in Sec.~\ref{sec:second-pro}.

\subsection{Two-color protocol}
\label{sec:first-pro}

The first protocol follows the idea of two-color, two-pulse adiabatic transfer proposed for single-spin states in \cite{chen04} and then further analyzed in Refs.~\cite{Caillet2007, Grodecka2007}. 
Here the system is driven by two two simultaneous laser pulses with frequencies
\begin{equation*}
    \omega_n = \left(\epsilon_X - \epsilon_n\right)/\hbar - \Delta,\quad n=T,S,
\end{equation*}
so that both pulses are detuned from the transition to the $\ket*{X}$ state by a common detuning $\Delta$ and their frequencies satisfy the Raman condition for the energy-conserving transfer between the states $\ket{S}$ and $\ket{T}$, as shown in  Fig.~\ref{fig:lambda-system}(a).
Correspondingly, we write 
\begin{equation*}
\bm{d}\cdot\bm{\mathcal{E}}^{(-)}(t)=\sum_{n=T,S}\Omega_n(t)e^{i(\omega_n t+\gamma_n)},
\end{equation*}
where $\Omega_n$ is assumed real.
The laser Hamiltonian defined by Eq.~\eqref{HL}, projected on the relevant subspace of system states, then takes the form
\begin{gather*}
    H_L = \dfrac{\hbar}{2} \sum_{n=T,S} \Omega_n(t) e^{i(\omega_n t + \gamma_n)} \frac{1}{\sqrt{2}}\left(\ket{T}+\ket{S}\right)\bra{X} + \mathrm{h.c.}
\end{gather*}

As only the difference between the phases of the laser pulses is relevant, we set $\gamma_S = 0$ and denote $\gamma_T=\gamma$. The pulses are assumed to have the same shape and arrive simultaneously; they are parametrized as 
\begin{gather*}
    \Omega_T(t) = \sqrt{2}\Omega(t) \cos\beta \quad \mathrm{and} \quad \Omega_S(t) = \sqrt{2} \Omega(t) \sin\beta.
\end{gather*}
We assume a Gaussian pulse shape 
$\Omega(t) = \Omega_0 \exp[-t^2/(2\tau^2)]$,
where $\tau$ is the pulse duration. It is convenient to introduce the further parametrization
\begin{equation*}
    \Omega(t) =\Theta(t) \sin\left[2\phi(t)\right],\quad
    \Delta    =\Theta(t) \cos\left[2\phi(t)\right],
\end{equation*}
i.e., 
\begin{equation*}
\Theta(t) = \sqrt{\Omega^2(t) + \Delta^2},  \quad \phi(t) = \frac{1}{2} \arctan \frac{\Omega(t)}{\Delta},
\end{equation*}
where $\phi(t)$ is referred to as the tipping angle.

We transform the system to the rotating frame picture using the unitary operator
\begin{equation}
\begin{split}
    U(t) =& \exp\left\{\frac{it}{\hbar}
    \left[\left(\epsilon_T-\epsilon_{X}+\hbar\Delta\right)\dyad*{T}\right.\right.\\
    &\left.\left.+ \left(\epsilon_S-\epsilon_X+\hbar\Delta\right)\dyad{S}{S}
    +\left(\epsilon_X-\hbar\Delta\right)\hat{\mathbbm{1}}\right]\right\}.
\end{split}
\label{eq:rotating-frame-unitary-operator}
\end{equation}
The Hamiltonian can then be split into two components,
\begin{gather}
    \tilde{H}_\mathrm{c} = \tilde{H}_\mathrm{c}^{(0)} + \tilde{H}_\mathrm{c}^{(1)}.
    \label{eq:general-form-of-hamiltonian-in-rotating-frame}
\end{gather}
The first term of Eq.~\eqref{eq:general-form-of-hamiltonian-in-rotating-frame} contains the secular terms,
\begin{gather}
    \tilde{H}_\mathrm{c}^{(0)} = \hbar\Delta \dyad{X} + \dfrac{\hbar}{2}\Omega(t) \left(\dyad{B}{X} + \dyad{X}{B}\right),
    \label{eq:hamiltonian-in-new-basis-dark-bright}
\end{gather}
where we introduce the bright state
\begin{gather}
    \ket{B} =  e^{i\gamma}\cos\beta\ket*{T} + \sin\beta\ket*{S}
    \label{eq:definition-of-bright-state}
\end{gather}
coupled to the $\ket*{X}$ state and the orthogonal dark state,
\begin{gather}
    \ket{D} = e^{i\gamma} \sin\beta \ket*{T} - \cos\beta \ket*{S},
    \label{eq:definition-of-dark-state}
\end{gather}
which is decoupled from the excited state and thus unaffected during the evolution.
The Hamiltonian $\tilde{H}_\mathrm{c}^{(0)}$ drives the intended evolution of the singlet-triplet qubit. 
The second part reads
\begin{equation}
\begin{split}
    \tilde{H}_\mathrm{c}^{(1)} =&  \dfrac{\hbar}{2}\Omega_T(t) e^{-2i\delta t+i\gamma} \frac{1}{\sqrt{2}} \dyad*{S}{X}\\
    &+\dfrac{\hbar}{2} \Omega_S(t) e^{2i\delta t} \frac{1}{\sqrt{2}}\dyad*{T}{X} + \mathrm{h.c.}
\end{split}
\label{eq:off-resonant-terms}
\end{equation}
This contribution contains off-resonant terms rotating with the frequency corresponding to the splitting between triplet and singlet states $2\hbar\delta = \epsilon_T - \epsilon_S$ that are treated as a perturbation. 

The essence of the protocol is to let the state $\ket*{B}$ undergo an adiabatic evolution generated by the Hamiltonian $\tilde{H}_\mathrm{c}^{(0)}$ and attain a dynamical phase due to the AC Stark shift induced by the laser field that off-resonantly couples this state to the excited state $\ket*{X}$.
This additional phase, relative to the decoupled dark state $\ket*{D}$, is equivalent to a rotation on the Bloch sphere around the axis defined by the states $\ket*{B}$ and $\ket*{D}$ which are, in turn, defined by the pulse parameters.
The adiabatic evolution is secured upon a sufficiently slow variation of $\Omega(t)$, by the splitting $\Delta$, which generates a gap between the instantaneous eigenvalues of $\tilde{H}_\mathrm{c}^{(0)}$ at any time $t$. In addition, the protocol relies on the spectral selectivity that suppresses off-resonant transitions induced by $\tilde{H}_\mathrm{c}^{(1)}$, for which a sufficiently long pulse duration (compared to the inverse exchange splitting) is required. 

To describe the ideal adiabatic evolution explicitly, we find the instantaneous eigenvalues of $\tilde{H}_\mathrm{c}^{(0)}$,
\begin{align*}
    &\lambda_0(t) = 0,\\
    &\lambda_1(t) = -\hbar\Theta(t)\sin^2\phi(t) = \dfrac{\hbar}{2} \left[\Delta - \sqrt{\Delta^2+\Omega^2(t)}\right],\\
    &\lambda_2(t) = \phantom{-}\hbar\Theta(t) \cos^2\phi(t) = \dfrac{\hbar}{2} \left[\Delta + \sqrt{\Delta^2 + \Omega^2(t)}\right],
\end{align*}
and the corresponding instantaneous eigenstates
\begin{align*}
    &\ket{a_0(t)} = \ket{D},\\
    &\ket{a_1(t)} = \cos\phi(t) \ket{B} - \sin\phi(t) \ket{X},\\
    &\ket{a_2(t)} = \sin\phi(t) \ket{B} + \cos\phi(t) \ket{X}.
\end{align*}
Before and after the pulse, the tipping angle satisfies $\phi(t\to\pm \infty)\to0$. 
Therefore, $\ket*{a_1(\pm\infty)}=\ket*{B}$ and $\ket*{a_2(\pm\infty)}=\ket*{X}$.
In general, the state of the system at time $t$ can be written as
\begin{gather}
    \ket{\psi(t)} = \sum_{n} c_n(t) e^{-i\Lambda_n(t)} \ket*{a_n(t)}.
    \label{eq:adiabatic_approximation}
\end{gather}
where $c_n(t)$ are time-dependent coefficients and 
\begin{equation}
    \Lambda_n(t) = \dfrac{1}{\hbar}\int_{t_0}^{t} \lambda_n(t')\dd t',
    \label{eq:Lambda}
\end{equation}
is the phase attained during the evolution.
By virtue of the adiabatic theorem \cite{Messiah_1966}, in the case of perfectly adiabatic evolution, the coefficients $c_n(t)$ are time-independent, thus the only effect of the evolution is the acquisition of a phase by each state.
The dark state remains intact while the bright and excited states receive an additional phase, $\Lambda_n\equiv\Lambda_n(\infty)$, where $n=1$ or $n=2$ refer to the bright and excited states, respectively.
The evolution (rotation) operator in the $\{\ket*{D},\ket*{B}, \ket*{X}\}$ basis is
\begin{gather*}
    U_\mathrm{c}(t) = \begin{pmatrix}
        1 & 0 & 0\\
        0 & e^{-i\Lambda_1(t)}\cos\phi(t) & e^{-i\Lambda_2(t)}\sin\phi(t)\\
        0 & -e^{-i\Lambda_1(t)}\sin\phi(t) & e^{-i\Lambda_2(t)}\cos\phi(t)
    \end{pmatrix}.
    \label{eq:full_rotation_operator}
\end{gather*}
After the pulse is switched off, $\phi(t)\to 0$ and the transformation, projected on the qubit subspace spanned by the states $\ket*{D}$ and $\ket*{B}$, has the simple form
\begin{equation*}
    U_\mathrm{c}^{(\mathrm{q})}(\infty) = 
    \dyad*{D}+e^{-i\Lambda_1}\dyad*{B}.
\end{equation*}
Converted to the original singlet-triplet basis according to Eq.~\eqref{eq:definition-of-bright-state} and Eq.~\eqref{eq:definition-of-dark-state},
this corresponds to the qubit rotation
\begin{gather*}
    U_\mathrm{c}^{(\mathrm{q})}(\infty) = \cos\dfrac{\Lambda_1}{2} \mathbbm{1} - i\sin\dfrac{\Lambda_1}{2}\vec{\bm{\sigma}}\cdot \vec{\mathbf{n}}
    = e^{-i\Lambda_1\vec{\bm{\sigma}}\cdot \vec{\bm{n}}/2},
\end{gather*}
where $\vec{\bm{\sigma}}$ is a vector of Pauli matrices expressed in the basis of $\{\ket{T},\ket{S}\}$.
The laser field thus induces a rotation on the Bloch sphere by an angle  $\Lambda_1$ 
about the axis given by
\begin{gather*}
    \vec{\mathbf{n}} = \left[\cos\gamma\sin\left(2\beta\right),-\sin\gamma\sin\left(2\beta\right),\cos\left(2\beta\right)\right].
    \label{eq:rotation_axis}
\end{gather*}
with the triplet and singlet placed at the poles along \makebox{$z$-axis}.
It is now clearly seen that by the appropriate choice of the laser amplitudes and phases, an arbitrary axis of the qubit rotation can be achieved. 
There is no unique relation between the pulse parameters and the rotation angle of the Bloch vector. 
In Fig.~\ref{fig:tipping_angle}(a) the scaled pulse area $\Omega_0/\Delta$ leading to a rotation angle of $\pi$ in the adiabatic evolution is plotted as a function of the scaled pulse duration. This shows that for any detuning and pulse duration, a pulse can be found which satisfies this condition.
Figure~\ref{fig:tipping_angle}(b) shows exemplarily the tipping angle and its derivative as a function of time for a $10$~ps pulse with a detuning of $1$~meV.

For future convenience, we express the initial state of the system as
\begin{gather*}
    \ket{\psi_0} = \cos\dfrac{\vartheta}{2}\ket{B} + e^{i\varphi}\sin\dfrac{\vartheta}{2}\ket{D},
\end{gather*}
where $\vartheta$ and $\varphi$ are the angles on a Bloch sphere.
We will also need two other states, orthogonal to the above
\begin{gather}
    \ket{\psi_1} = \sin\dfrac{\vartheta}{2}\ket{B} - e^{i\varphi}\cos\dfrac{\vartheta}{2}\ket{D},\quad \ket{\psi_2} = \ket{X}.
    \label{eq:stany-ortogonalne-do-psi-zero}
\end{gather}

\subsection{Single-color protocol}
\label{sec:second-pro}

The second protocol, implemented experimentally in Ref.~\cite{Kim2011}, uses only a single pulse, centered between the energies of the singlet and triplet states and detuned from the $\ket*{X}$ state by $\hbar\Delta$ (see Fig.~\ref{fig:lambda-system}(b)). This reduces the experimental complexity but, as we will see, restricts the qubit rotation to a single axis. The essential mechanism of the qubit rotation is still the phase accumulation attained during the adiabatic evolution along an AC-Stark-shifted spectral branch, but the different spectral conditions lead to a different nature of the corrections to the ideal evolution. 

The Hamiltonian describing the laser driving now takes the form
\begin{equation*}
H_L = \frac{\hbar}{2}\Omega(t)e^{i \omega t} \frac{1}{\sqrt{2}}\left(\ket{T} + \ket{S}\right)\bra*{X} + \textrm{h.c.}
\label{eq:hamiltonian-single-pulse}
\end{equation*}
Using the same unitary transformation of Eq.~\eqref{eq:rotating-frame-unitary-operator} as previously, we again transform the Hamiltonian to the rotating frame.
The new Hamiltonian is again split as in Eq.~\eqref{eq:general-form-of-hamiltonian-in-rotating-frame} but now we define
\begin{align}
\begin{aligned}
    \tilde{H}_\mathrm{c}^{(0)} = &\hbar\Delta \dyad*{X}
    \\& + \frac{\hbar}{2}\Omega(t) \frac{1}{\sqrt{2}}\left(\ket*{T}+\ket*{S}\right)\bra*{X} + \textrm{h.c.}
\end{aligned}
\label{H0-1p}
\end{align}
and
\begin{align*}
\begin{aligned}
    \tilde{H}_\mathrm{c}^{(1)} = \frac{\hbar}{2}\Omega(t) \frac{1}{\sqrt{2}} \left[ \left(e^{i\delta t}-1\right) \ket*{T}\right.\\
    \left.+ \left(e^{-i\delta t}-1\right) \ket*{S}\right]\bra*{X} + \textrm{h.c.}
\end{aligned}
\end{align*}

As follows from Eq.~\eqref{H0-1p}, the bright state is now
\begin{gather*}
    \ket*{B} = \frac{1}{\sqrt{2}} \left( \ket*{T} + \ket*{S} \right)
    \label{eq:bright-second-protocol}
\end{gather*}
and the dark state is
\begin{gather*}
    \ket*{D} = \frac{1}{\sqrt{2}} \left( \ket*{T} - \ket*{S} \right).
    \label{eq:dark-second-protocol}
\end{gather*}
With these definitions of the states, the Hamiltonian $\tilde{H}_\mathrm{c}^{(0)}$ is exactly the same as in the case of the two-color protocol [Eq.~\eqref{eq:hamiltonian-in-new-basis-dark-bright}]. However, the
 qubit rotation is now performed only about the $x$-axis on the Bloch sphere, corresponding to the transformation
\begin{gather*}
    U_\mathrm{c}^{(\mathrm{q})}(\infty) = \cos\frac{\Lambda_1}{2} \mathbbm{1} - i \sin\frac{\Lambda_1}{2} \sigma_x.
\end{gather*}

\section{Imperfections of the unitary evolution
\label{sec:leakage-of-quantum-information}}

\begin{figure}[t]
    \centering
    \includegraphics[width=\linewidth]{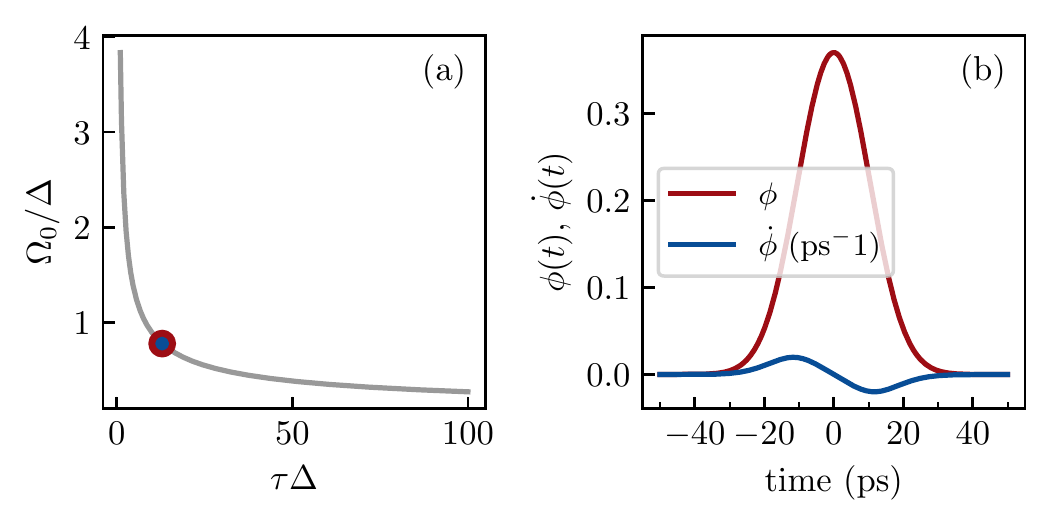}
    \caption{ (a) Scaled pulse area $\Omega_0/\Delta$ as a function of the scaled pulse duration $\tau\Delta$ corresponding to the rotation over angle of $\Lambda_1 = \pi$ rad.
    Red-blue point corresponds to the value of detuning and pulse duration for which $\phi(t)$ and $\dot{\phi}(t)$ were depicted in (b). (b) Tipping angle $\phi(t)$ (red line) and its derivative~$\dot{\phi}(t)$ (blue line) as a function of time for pulse duration $\tau=10$~ps and detuning $\hbar\Delta=1$~meV.}
    \label{fig:tipping_angle}
\end{figure}

In this section, we analyze the discrepancies between the idealized qubit rotation based on the adiabatic theorem and the actual system evolution.
To the lowest order, the imperfections can be divided in two classes. First, non-adiabatic corrections are revealed as the difference between the exact evolution generated by $ \tilde{H}_\mathrm{c}^{(0)}$ and the adiabatic approximation. Second, the off-resonant terms contained in $\tilde{H}_\mathrm{c}^{(1)}$ may affect the evolution. We study the magnitude of these corrections for a fixed intended angle of qubit rotation of $\Lambda_1(\infty)=\pi$ around the $x$ axis in both protocols. That is, for each set of parameters (exchange splitting, detuning, and pulse duration) we determine $\Omega_0$ from Eq.~(\ref{eq:Lambda}) with $\Lambda_1(\infty) = \pi$, as shown in Fig.~\ref{fig:tipping_angle}(a).
One should note that the  corrections discussed here do not involve dephasing and can in principle be compensated for by an appropriate adjustment of the control fields. 

\subsection{Non-adiabatic transitions}

\begin{figure}[t]
    \centering
    \includegraphics[width=\linewidth]{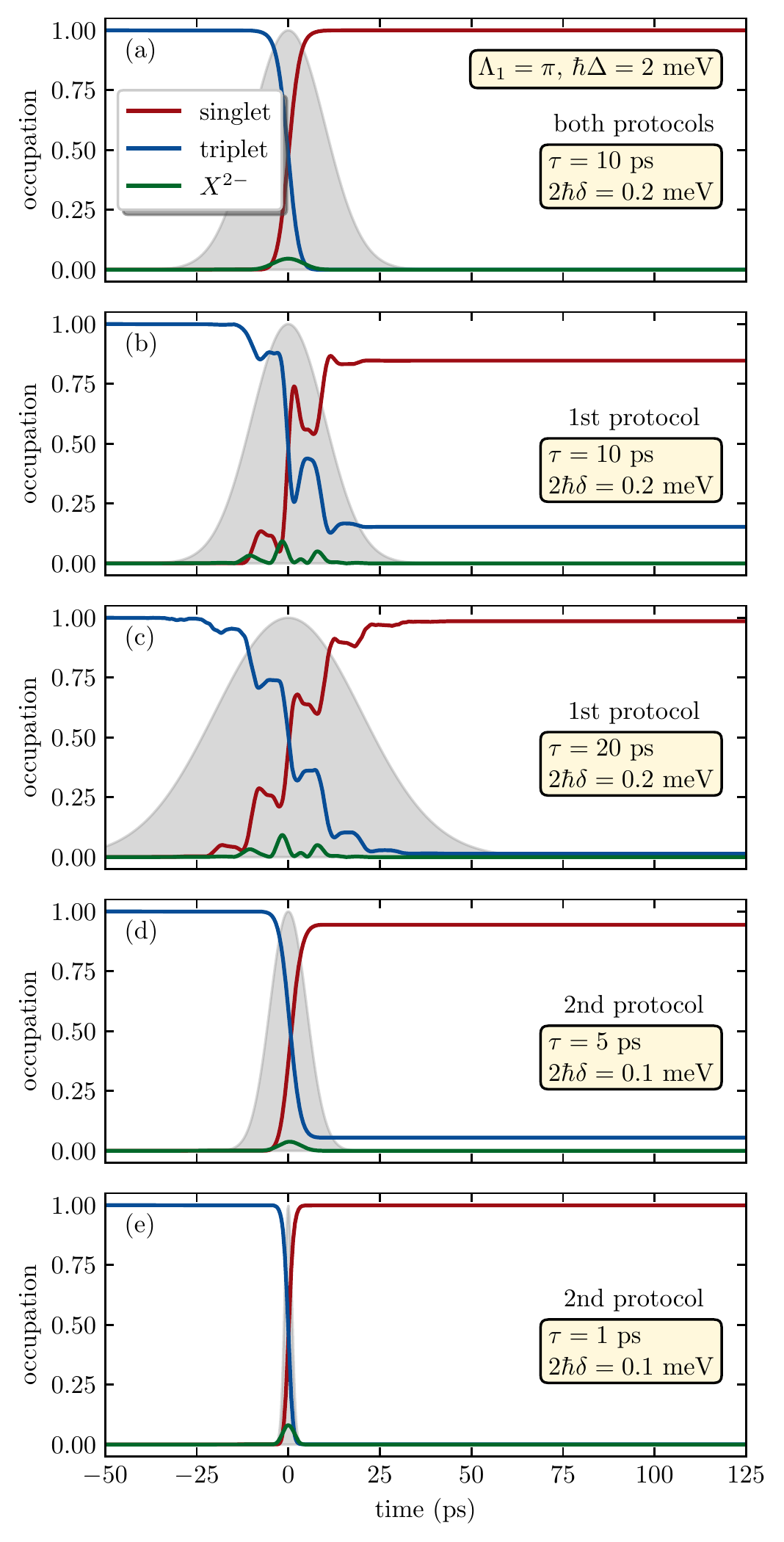}
    \caption{Occupancy of singlet, triplet and $X^{2-}$ states in time during gating operations within the first protocol (a) when evolution is governed by a Hamiltonian $H_\mathrm{c}^{(0)}$; (b-e) when evolution is governed by a Hamiltonian $H_\mathrm{c} = H_\mathrm{c}^{(0)} + H_\mathrm{c}^{(1)}$, i.e., containing off-resonance terms in the first (b,c) and second (d,e) type of protocol for different values of pulse duration and singlet-triplet energy splitting.}
    \label{fig:evolution-first-protocol}
\end{figure}

An example of the exact (numerically computed) evolution governed by $ \tilde{H}_\mathrm{c}^{(0)}$ and starting from the triplet state is shown in Fig.~\ref{fig:evolution-first-protocol}(a).
For these particular parameters, the adiabatic approximation works very well.
The numerically calculated evolution agrees almost perfectly with the one obtained within the adiabatic scheme and the $\pi$ rotation to the singlet state is achieved with a very high fidelity.
Figure \ref{fig:evolution-first-protocol}(b) and (c) show results for the evolution of the occupations of the states obtained from a numerical simulation of the two-color protocol including the full carrier-light Hamiltonian $H_\mathrm{c}=H_\mathrm{c}^{(0)}+H_\mathrm{c}^{(1)}$ for a 10~ps and a 20~ps pulse. We find a clear influence of the off-resonant part during the presence of the pulse and a reduction of the fidelity in the case of the 10~ps pulse.
With increasing pulse duration, the fidelity increases, which can be understood from the increased spectral selectivity of the longer pulse.
Figure~\ref{fig:evolution-first-protocol}(d) and (e) shows the evolution of the occupations in the case of the single-color protocol for a 5~ps and a 1~ps pulse.
In contrast to the first protocol, now the fidelity increases with decreasing pulse duration because of the increasing simultaneous overlap with both transitions.
To perform a more complete and quantitative analysis, we choose the final occupation of the auxiliary state $\ket*{X}$ as a figure of merit. Such an occupation may result from a non-adiabatic jump between the branches of instantaneous eigenvalues and constitutes a ``leakage'' error, as the occupation leaves the computational singlet-triplet space. 

We find an approximate formula for the final occupation of state $\ket*{X}$ from the lowest-order correction to the adiabatic evolution \cite{Messiah_1966}.
Substituting Eq.~\eqref{eq:adiabatic_approximation} to the Schr\"odinger equation one can easily derive the equation for the coefficients $c_n(t)$,
\begin{gather*}
    \dot{c}_m(t) = -\sum_n c_n(t) \braket{a_m(t)}{\dot{a}_n(t)}.
    \label{eq:equation-adibatic-theorem}
\end{gather*}
We aim at obtaining an approximate formula for the amplitude $c_2(\infty)$ which corresponds to the occupation of the $\ket*{X}$ state at the end of the gating procedure.
For $m=2$ the only relevant term on the right-hand side is $\braket*{a_2(t)}{\dot{a}_1(t)} = -\dot{\phi}(t)$, since $\braket{a_n(t)}{\dot{a}_n(t)} = 0$ for any $n$ and $\ket*{a_0(t)}$ is time independent. 
Finally, we take the upper estimate of the integral by setting $c_1(t) = c_1(0) = \cos(\vartheta/2)$ and write
\begin{gather}
    c_2^\mathrm{na} = \cos\frac{\vartheta}{2}\int_{-\infty}^\infty \dd  t e^{i\left[\Lambda_2(t)- \Lambda_1(t)\right]} \dot{\phi}(t)
    \label{eq:c_2_na_amplitude}
 \end{gather}
 where the superscript ``na'' corresponds to the non-adiabatic corrections.
Thus, the error due to non-adiabacity is
\begin{gather}
    \delta_\mathrm{u}^{\mathrm{na}} = \left|c_2^{\mathrm{na}}\right|^2.
    \label{eq:delta-na}
\end{gather}
The integral in Eq.~\eqref{eq:c_2_na_amplitude} can be evaluated as
\begin{align*}
\begin{aligned}
    c_2^\mathrm{na} &= -i\cos\frac{\vartheta}{2} \tau\Delta \tilde{\Omega}_0\\
    &\times \int_0^\infty \dd s \sin\left(\Lambda_1 \frac{F(\tilde{\Omega}_0,s) +s }{F(\tilde{\Omega}_0,\infty)}\right)\frac{s e^{-s^2/2}}{1+\tilde{\Omega}_0^2 e^{-s^2}}
\end{aligned}
\end{align*}
where we have performed the change of integration variables $s=t/\tau$ and defined $\tilde{\Omega}_0 = \Omega_0/\Delta$ together with the function
\begin{gather*}
    F(\tilde{\Omega}_0,s) = \int_0^s \dd s' \left(\sqrt{1 + \tilde{\Omega}_0 e^{-s'^2}}-1\right).
\end{gather*}
We benefit from the oddness of $F(\tilde{\Omega}_0,s)$ and $\dot{\phi}(t)$ and use the relation
\begin{gather*}
    \Lambda_1 = \tau \Delta F( \tilde{\Omega}_0, \infty)
\end{gather*}
It is now clearly seen that $\delta_\mathrm{u}^\mathrm{na}$ depends only on the product of $\tau$ and $\Delta$.

This estimate for $\delta_\mathrm{u}^\mathrm{na}$ is shown in Fig.~\ref{fig:na+off-resonant}(a,b) as a function of $\tau$ and $\Delta$. As expected, the error very quickly drops down by orders of magnitude when $\tau\Delta \gtrsim 1$. The oscillations visible in the plots are due to the interference of quantum amplitudes for non-adiabatic jumps on the rising and decreasing slopes of the pulse, where the rate of change of the tipping angle is the fastest (see Fig.~\ref{fig:tipping_angle}(b)) \cite{Grodecka2007}. 

\subsection{Off-resonant terms: two-color protocol}

\begin{figure}[t]
    \centering
    \includegraphics[width=\linewidth]{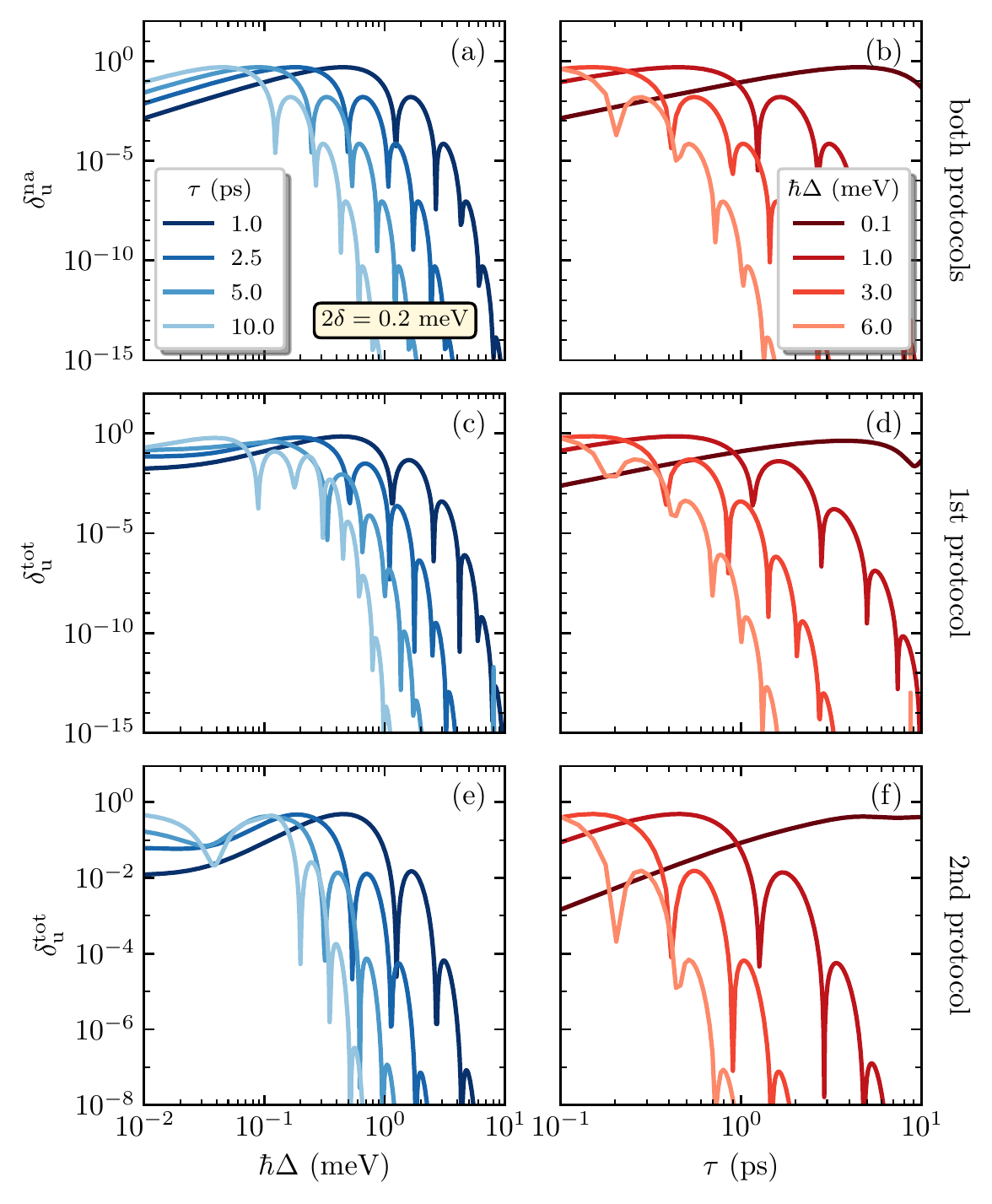}
    \caption{Leakage to the excited state within the first scheme of optical control for the triplet-singlet splitting $2\hbar\delta = 0.2$ meV and averaged over all possible initial states on a Bloch sphere: (a,b,c) as a function of the detuning for three different values of the pulse duration; (b,d,f) as a function of pulse duration for several values of the detuning.}
    \label{fig:na+off-resonant}
\end{figure}

The other source of corrections are the transitions induced by the Hamiltonian $\tilde{H}_\mathrm{c}^{(1)}$, which can be written in the bright-dark basis in the form
\begin{align*}
\begin{aligned}
\tilde{H}_c^{(1)} =& \frac{\hbar}{2}\Omega(t) \left[\sin^2\beta e^{i\left(2\delta t-\gamma\right)} - \cos^2\beta e^{-i\left(2\delta t-\gamma\right)}\right]\dyad*{D}{X}\\
&+ \frac{\hbar}{2}\Omega(t) \sin(2\beta) \cos\left(2\delta t - \gamma\right) \dyad*{B}{X} + \textrm{h.c.}
\end{aligned}
\end{align*}
First, we study the unwanted occupation of the state $\ket*{X}$, which is found by
treating $\tilde{H}_c^{(1)}$ as a perturbation to the ideal adiabatic evolution.
Using first-order perturbation theory, we obtain the relevant probability amplitude,
\stepcounter{equation}
\begin{align*}
     c_2^\mathrm{off} =&\frac{1}{\hbar}\int\limits_{-\infty}^\infty \dd t \mel*{X}{U_c^\dagger(t) \tilde{H}_c^{(1)} U_c(t)}{\psi_0} \\
     = & \frac{1}{2}\cos\frac{\vartheta}{2}\sin\left(2\beta\right)\int_{-\infty}^\infty \dd t e^{i\left[\Lambda_2(t)-\Lambda_1(t)\right]}\Omega(t)\\
     &\times\cos\left(2\delta t - \gamma\right)\cos\left[2\phi(t)\right]\\
     &+\frac{1}{2}e^{i\varphi} \sin\frac{\vartheta}{2} \int_{-\infty}^\infty \dd t e^{i\Lambda_2(t)}\Omega(t) \cos\phi(t)\\
     &\times\left[e^{-i\left(2\delta t-\gamma\right)}\sin^2\beta - e^{i\left(2\delta t-\gamma\right)}\cos^2\beta\right]. \tag{\theequation}
    \label{eq:off-resonant-amplitude-probab-first-protocol}
\end{align*}
The corresponding error is defined as
\begin{gather*}
    \delta_\mathrm{u}^\mathrm{off} = |c_2^\mathrm{off}|^2.
\end{gather*}
The total leakage of quantum information into the excited state, taking into account the possible interference between the two mechanisms of leakage, is
\begin{gather}
    \delta_\mathrm{u}^\mathrm{tot} = |c_2^\mathrm{na} + c_2^\mathrm{off}|^2
    \label{eq:delta-tot}
\end{gather}
This total leakage probability, averaged over the initial state, is shown in Fig.~\ref{fig:na+off-resonant}(c,d) as a function of $\tau$ and $\Delta$. It turns out that including the effects of $\tilde{H}_\mathrm{c}^{(1)}$ has a minor effect on the leakage probability and the non-adiabatic jumps remain the main source of this kind of errors.

The second type of error induced by $ \tilde{H}_c^{(1)}$ is the modification of the final state within the qubit space. This is quantified by the probability of finding the system in the state orthogonal to the intended one. In the leading order, for an initial state $\ket*{\psi_{0}}$ this is given by 
$\delta_{T\text{-}S}=|c_{T\textrm{-}S}|^2$, where
\begin{equation*}
c_{T\text{-}S} = -\frac{i}{\hbar}\int\limits_{-\infty}^\infty \mel{\psi^\bot}{U_c^\dagger(t) \tilde{H}_c^{(1)} U_c(t)}{\psi_{0}}\dd t,
\end{equation*}
where $\ket*{\psi^\bot}$ is the qubit state orthogonal to $\ket*{\psi_{0}}$.
Restricting ourselves to $\ket*{\psi_{0}}=\ket*{T}$ we get
\begin{align}
c_{T\text{-}S} =& \frac{1}{\hbar}\int\limits_{-\infty}^\infty \mel{S}{U_c^\dagger(t) \tilde{H}_c^{(1)} U_c(t)}{T}\dd t \nonumber\\
=&\frac{1}{2}e^{-i\gamma}\int\limits_{-\infty}^\infty \dd t \Omega(t) \sin{\left[\phi(t) \right]} \left\{ \cos^{2}{\left(\beta \right)} e^{- i \Lambda_{1}(t)} \right.\nonumber\\
&\times \left[e^{i \left(2\delta t - \gamma\right)} \sin^{2}{\left(\beta \right)} - e^{- i \left(2\delta t - \gamma\right)} \cos^{2}{\left(\beta \right)}\right] \nonumber\\
&-\sin{\left[\phi(t)\right]} e^{i \Lambda_{1}(t)} \sin^{2}{\left(\beta \right)}\nonumber\\
&\times \left[- e^{i \left(2\delta t - \gamma\right)} \cos^{2}{\left(\beta \right)} + e^{- i \left(2\delta t - \gamma\right)} \sin^{2}{\left(\beta \right)}\right] \nonumber\\
&\left.-\sin^{2}{\left(2 \beta \right)} \cos{\left[\phi(t) \right]} \cos{\left(2\delta t - \gamma \right)} \right\}.
\label{eq:off-resonant-influence-on-qubit-rotation-two-pulses}
\end{align}
This contribution to the error is  plotted in Fig.~\ref{fig:comparison_of_protocols}(a,b) as a function of pulse duration and exchange splitting.
It is clear that the fidelity grows for longer pulses and larger exchange splitting. Unlike the leakage to the $\ket*{X}$ state, this error is due to imperfect spectral selectivity of the singlet and triplet states by the pulses deriving the $\Lambda$-system and therefore it depends on the exchange splitting, which is much lower than typically achievable detunings from the excited state. As a result, this error dominates in the range of longer pulse durations, where the leakage (cf. Fig.~\ref{fig:na+off-resonant}(c,d)) is suppressed.

\subsection{Off-resonant terms: single-color protocol}
\begin{figure}[tb]
    \centering
    \includegraphics[width=\linewidth]{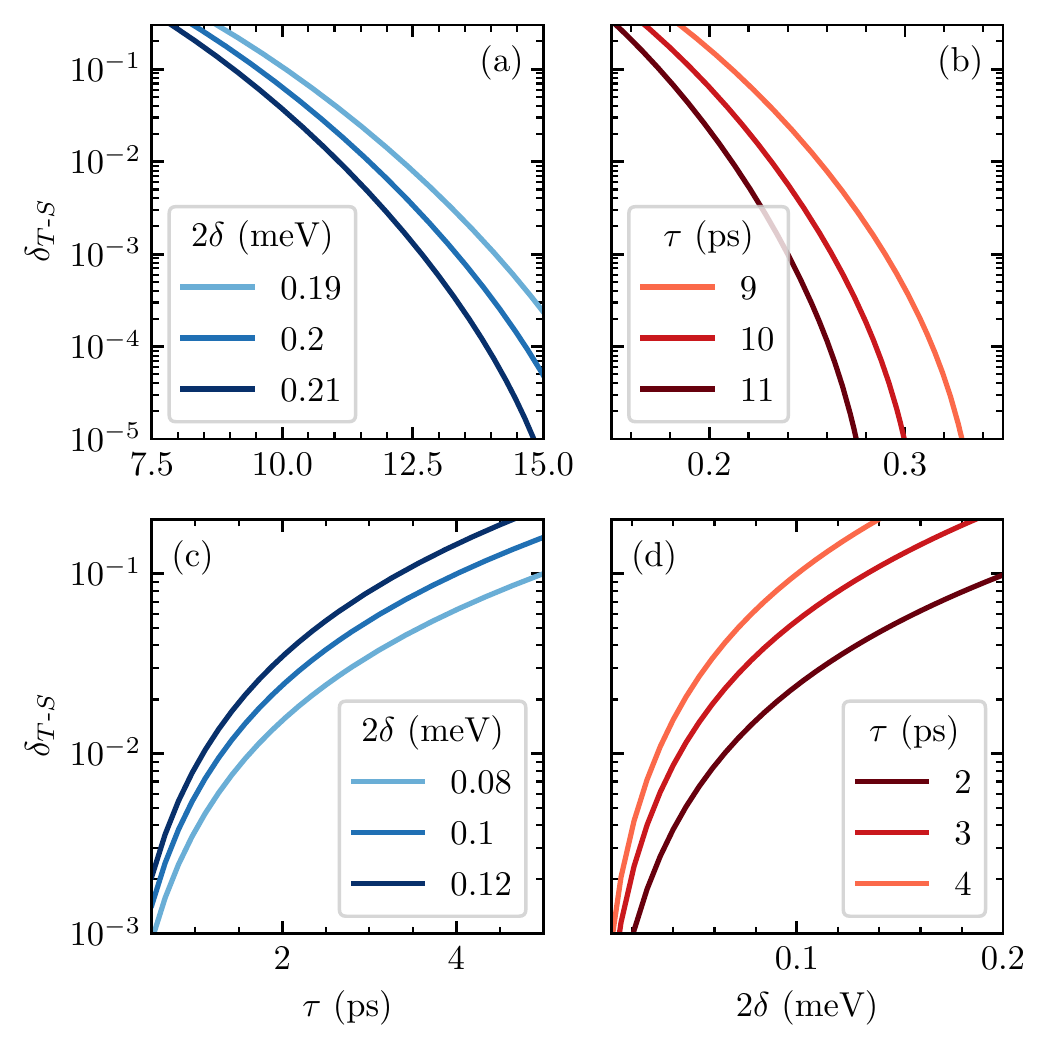}
    \caption{Quantum gate error due to the off-resonant terms in $H_\mathrm{c}$ within the first (a,b) and second (c,d) type of protocol as a function of pulse duration (a,c) and singlet-triplet energetic splitting (b,d).}
    \label{fig:comparison_of_protocols}
\end{figure}
The off-resonant part in the second type of protocol in the basis of bright and dark states takes form
\begin{align*}
\tilde{H}_\mathrm{c}^{(1)} =&\frac{\hbar}{2}\Omega(t) \Bigl\{\left[\cos\left(\delta t\right) - 1\right]\ket*{B}\\
&+\left.\sin\left(\delta t\right)\ket*{D}\right]\Bigl\}\bra*{X} +\textrm{h.c.}
\label{eq:off-resonant-terms-second-protocol}
\end{align*}
As in Eq.~\eqref{eq:off-resonant-amplitude-probab-first-protocol}, we can write the leakage amplitude $c_2^{\mathrm{off}}$ for an arbitrary initial state in the leading order as
\begin{align*}
    c_2^\mathrm{off} =& \frac{1}{2}\cos\frac{\vartheta}{2} \int\limits_{-\infty}^\infty \dd t \Omega(t)e^{i[\Lambda_2(t)-\Lambda_1(t)]}\\
    &\qquad\times \cos\left[2\phi(t)\right]\left[\cos\left(\delta t\right)-1\right]\\
    &+\frac{1}{2}e^{i\varphi}\sin\frac{\vartheta}{2}\int\limits_{-\infty}^\infty \dd t \Omega(t) e^{i\Lambda_2(t)}\cos\phi(t) \sin\left(\delta t\right)
\end{align*}
This total leakage error for the single-pulse protocol, averaged over the initial states, is plotted in Fig.~\ref{fig:na+off-resonant}(e,f). Again, the leakage induced by non-adiabaticity dominates. 

The amplitude of the error component within the qubit subspace for the initial trion state for this protocol is (cf. Eq.~\eqref{eq:off-resonant-influence-on-qubit-rotation-two-pulses})
\begin{align*}
c_{T\text{-}S} =&-\frac{1}{4} \int\limits_{-\infty}^\infty \dd t \Omega(t) e^{-i\Lambda_1(t)} \sin\left[\phi(t)\right]\\
&\times \left\{\left[2\vphantom{e^{\Lambda}}(\cos(\delta t) - 1)\cos\left[\phi(t)\right]
+  e^{i\Lambda_{1}(t)}\sin\left(\delta t\right)\right]e^{i\Lambda_{1}(t)}\right.\\ &\left. \vphantom{e^\Lambda}- \sin(\delta t)\right\}.
\end{align*}
The corresponding error $\delta_{T\text{-}S}=| c_{S\text{-}T}|^2$ is plotted in Fig.~\ref{fig:comparison_of_protocols}(c,d). 
In contrast to the previous case, the fidelity decreases with increasing pulse length and increasing singlet-triplet energy splitting. This is related to the construction of this protocol, which relies on the simultaneous coupling of both the singlet and triplet states to the excited state. 
As the pulse duration increases, the pulse becomes energetically narrow, thus unable to cover both qubit states.
Similarly, as the singlet-triplet splitting increases, it becomes more difficult to cover both states by a pulse of a given spectral width.
This can also be clearly seen from the form of the off-resonance terms.
Small $\delta$ assures $\sin(\delta\tau)$ and $\cos(\delta\tau)-1$ to be small within the given time scale.
On the other hand, as the time scale increases (larger $\tau$), a smaller value of delta is needed to keep the terms of $H_\mathrm{c}^{(1)}$ small.

While for the two-pulse protocol extending the pulse duration suppresses all types of errors discussed so far, in the single-pulse scheme, the error $\delta_{T\text{-}S}$ is reduced by decreasing the pulse duration, which leads to a trade-off with the leakage. Considerable improvement of the achievable fidelity is possible for smaller exchange splittings.

\section{Environment-induced errors}
\label{sec:env}

In this section, we study the effect of the lattice and radiative environments on the operation of the singlet-triplet qubit in the two control schemes. Unlike the unitary corrections discussed above, the decoherence induced by the interaction with the macroscopic surroundings of the qubit is irreversible and cannot be compensated by small adjustments of the control parameters. 

\subsection{General theory \label{sec:general-theory of-error}}
In this subsection, we give the most important steps in deriving the effect of decoherence due to the interaction with the environment on the evolution of a quantum system. This approach follows the general perturbative scheme \cite{Alicki2002,Roszak2005} that is valid whenever the overall environment-induced correction to the system state is small. 
The general form of the interaction Hamiltonian between the system and its environment is
\begin{equation}
    V = \sum_{l} \widehat{S}_{l} \otimes \widehat{R}_{l},
    \label{eq:general-form-of-interaction}
\end{equation}
where $\widehat{S}_{l}$ act in the Hilbert space of the system and $\widehat{R}_{l}$
act on the space of the environment.
The state of the system together with the environment is described by a density matrix
$\varrho(t)$.
The system is assumed to be initially in a product state 
$\varrho(t_0) = \rho(t_0)\otimes\rho_{\mathrm{env}}$,
with the system in a pure state, $\rho(t_0) \equiv \rho_0 = \dyad*{\psi_0}$, while the environment is taken to remain in thermal equilibrium.
One assumes that the evolution of the system and the environment in the absence of the coupling is known and given by
\begin{equation*}
    U_0(t) = U_\mathrm{c}(t) \otimes e^{- i t H_\mathrm{env} /\hbar},
\end{equation*}
where the system evolves according to $U_\mathrm{c}(t)$ and $H_\mathrm{env}$ is the Hamiltonian of the environment.
In the second-order Born approximation and in the interaction picture with respect to the unperturbed evolution $U_0(t)$, the equation for the density matrix reads
\begin{gather}
\begin{split}
    \widetilde{\varrho}\left(t\right) =& \widetilde{\varrho}\left(t_0\right) + \frac{1}{i\hbar} \int\limits_{t_0}^t \dd \tau \left[V(\tau), \varrho(t_0)\right]\\
    &-\dfrac{1}{\hbar^2}\int_{t_0}^t \dd  \tau \int\limits_{t_0}^\tau \dd \tau' \left[V(\tau), \left[V(\tau'),\varrho(t_0)\right]\right],
\end{split}
\label{eq:rownanie-na-macierz-gestosci-przyblizenie-Borna}
\end{gather}
where
$\widetilde{\varrho}(t) = U_0^\dagger(t) \varrho(t) U_0(t)$ and $\quad V(t) = U_0^\dagger(t) V U_0(t)$.
From the above, one extracts the reduced density matrix of the carrier subsystem (in the Schrödinger picture),
\begin{gather*}
    \rho(t) = U_\mathrm{c}(t)
     \Tr_R \widetilde{\varrho}(t)U_\mathrm{c}^{\dagger}(t),  
\end{gather*}
where the trace is taken over the reservoir degrees of freedom. The first (0th order) term in Eq.~\eqref{eq:rownanie-na-macierz-gestosci-przyblizenie-Borna} yields
$\rho^{(0)}(t) = U_\mathrm{c}(t)\rho_0 U_\mathrm{c}^\dagger(t) $.
The second (1st order) term vanishes because it contains an average of an odd number of reservoir operators which is zero in the thermal equilibrium state.
The third (2nd order) term is the leading correction to the dynamics of the system,
\begin{gather}
    \widetilde{\rho}^{(2)}(t) = -\dfrac{1}{\hbar^2} \int\limits_{t_0}^t \dd \tau \int\limits_{t_0}^\tau \dd \tau' \Tr_R\left[V(\tau), \left[V(\tau'),\varrho(t_0)\right]\right].
    \label{eq:second-order-rho-density-matrix-to-error}
\end{gather}
Then one can write
\begin{gather*}
    \rho(t) = U_\mathrm{c}(t)\left[\rho_0 +\widetilde{\rho}^{(2)}(t)\right]U_c^\dagger(t).
\end{gather*}
The fidelity $F$ of the gating procedure is defined as the overlap of the density matrix with the ideal final state $U_\mathrm{c}(t)\ket*{\psi_0}$,
\begin{gather}
    F^2 = \mel*{\psi_0}{U_\mathrm{c}(t)^\dagger \rho(t) U_\mathrm{c}(t) }{\psi_0} = 1 - \delta
    \label{eq:fidelity-definition}
\end{gather}
where
\begin{gather}
    \delta = \mel*{\psi_0}{\rho^{(2)}(t)}{\psi_0}
    \label{eq:delta-error-general-definition}
\end{gather}
is the error of the gating procedure.
Using Eqs.~\eqref{eq:fidelity-definition}, \eqref{eq:delta-error-general-definition} and \eqref{eq:second-order-rho-density-matrix-to-error} one can obtain the error of the quantum evolution due to the interaction with the environment,
\begin{gather}
    \delta = \sum_{ll'}\int \dd \omega R_{ll'}(\omega) S_{ll'}(\omega) ,
    \label{eq:error-environment}
\end{gather}
where we introduce two spectral functions: the spectral density of the reservoir
\begin{equation}
    R_{ll'}(\omega) = \frac{1}{2\pi}\int \dd t e^{i\omega t} \left\langle \widehat{R}_{l}(t) \widehat{R}_{l'}\right\rangle 
    \label{eq:definition-of-spectral-density-general-form}
\end{equation}
and the spectral characteristic function of the system evolution
\begin{equation}
    S_{ll'}(\omega) = \sum_i \mel*{\psi_0}{Y_{l}^\dag(-\omega)}{\psi_i} \mel*{\psi_i}{Y_{l'}(\omega)}{\psi_0}
    \label{eq:spectral-characteristic-definintion-general-form}
\end{equation}
where the sum is over states orthogonal to $\ket*{\psi_0}$ and
\begin{equation}
    Y_{l}(\omega) = \int \dd t e^{i\omega t} \widehat{S}_{l}(t),
    \label{eq:error-y-function}
\end{equation}
where $\widehat{S}_{l}(t) = U_c^\dagger(t)\widehat{S}_{l} U_c(t)$ are the system operators in the interaction picture.

\begin{figure}[t]
    \centering
    \includegraphics[width=\linewidth]{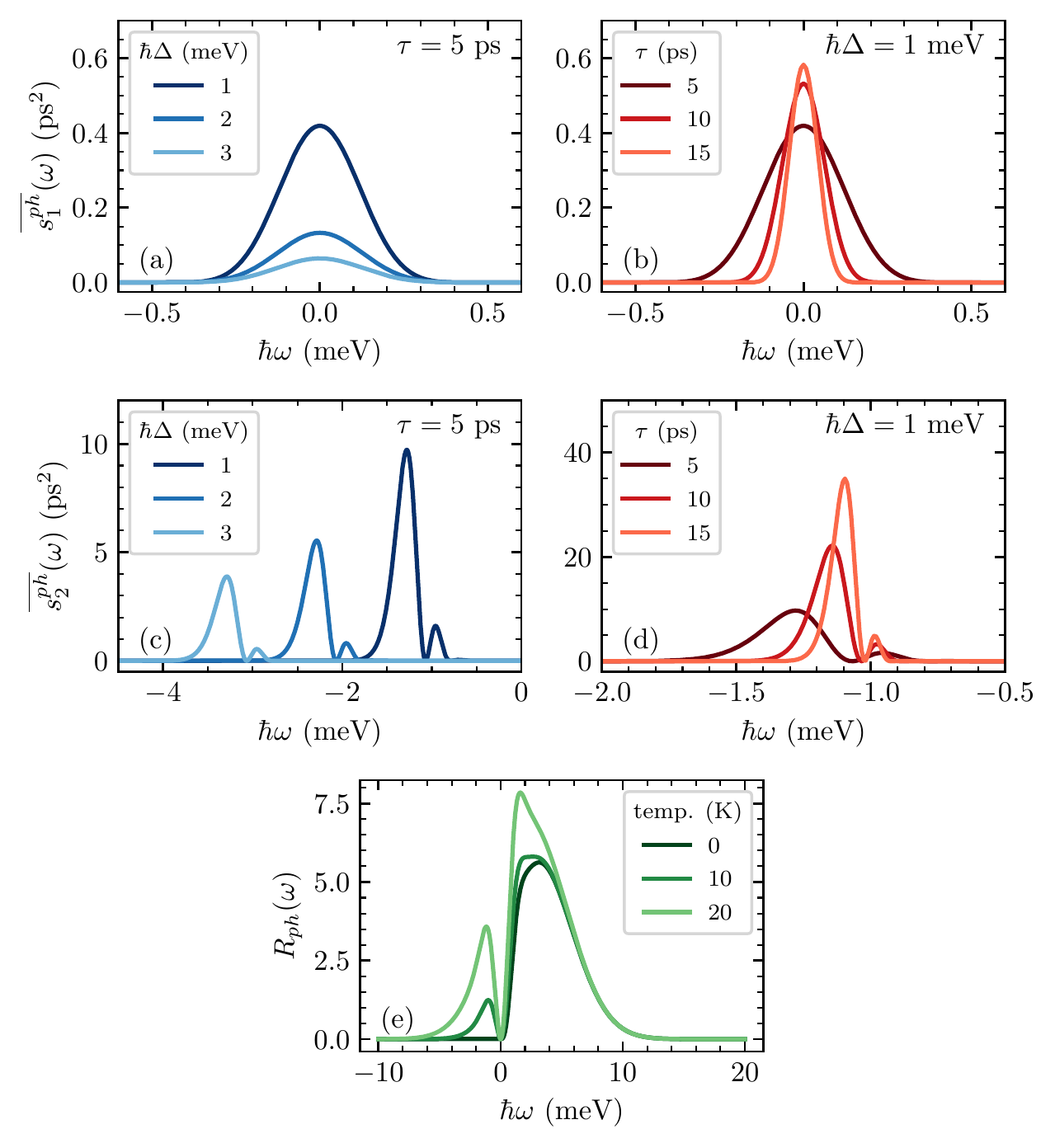}
    \caption{Spectral characteristics: (a,b) $\overline{s_1^\mathrm{ph}}(\omega)$ (see Eq.~\eqref{eq:average_spec_char_1}) and (c,d) $\overline{s_2^\mathrm{ph}}(\omega)$ (see Eq.~\eqref{eq:average_spec_char_2}) as a function of the phonon frequency, for different values of the the detuning at a fixed pulse length (a,c) and various values of the pulse duration at a fixed detuning (b,d). 
    (e) Spectral density (see Eq.~\eqref{eq:phonons_spectral_density}) as a function of the phonon frequency for three selected values of temperature.}
    \label{fig:spectral_characteristics}
\end{figure}

\subsection{Destructive influence of the phonon bath \label{sec:phonon-bath}}
In this subsection, we derive the gating error due to carrier-phonon interactions.
For that purpose, we employ the general theory of decoherence presented in~Sec.~\ref{sec:general-theory of-error} to the carrier-phonon Hamiltonian given by Eq.~\eqref{eq:phonon-carrier-interaction-simpified}.
The results presented here are very similar to those of Ref.~\cite{Grodecka2007} due to formal equivalence of the two Hamiltonians. 

Comparing Eq.~\eqref{eq:phonon-carrier-interaction-simpified} with Eq.~\eqref{eq:general-form-of-interaction} one can see that the system is coupled to phonons only through the diagonal coupling to the excited state, hence the only relevant operators are
\begin{equation*}
    \widehat{R}_{\mathrm{ph}} =  \sum_\mathbf{k} F(\mathbf{k})\left(\tilde{b}_\mathbf{k} + \tilde{b}_{-\mathbf{k}}^\dagger\right)
\end{equation*}
and $\widehat{S}_{\mathrm{ph}} = \dyad*{X}$.

Using Eq.~\eqref{eq:definition-of-spectral-density-general-form} and the standard properties of the bosonic thermal bath, one obtains the spectral density in the form
\begin{align}
\begin{aligned}
    R_\mathrm{ph}(\omega) =& \left[n_B(\omega,T) + 1\right]\\
    &\times \sum_\mathbf{k} |F\left(\mathbf{k}\right)|^2 \left[\delta(\omega-\omega_\mathbf{k}) + \delta(\omega+\omega_\mathbf{k})\right]
\end{aligned}
    \label{eq:phonons_spectral_density}
\end{align}
where $n_B(\omega)$ stands for the Bose-Einstein distribution at some temperature $T$.
The sum can be evaluated by approximating it by an integral and representing $\mathbf{k}$ in the spherical coordinates $\mathbf{k} = (k,\zeta,\eta)$.
The result is
\begin{equation}
    R_\textrm{ph}(\omega) = R_0 \left[n_B(\omega) + 1 \right] \omega^3 g(\omega),
\end{equation}
with $R_0 = (D_\mathrm{e} - D_\mathrm{h})/(8 \pi^2 \hbar \rho c_l^5)$
and
\begin{equation*}
    g(\omega) = \int\limits_{-\pi/2}^{\pi/2} \dd \zeta \cos\zeta \exp \left[ -\frac{\omega^2}{c_l^2} \left( l^2\sin^2\zeta + l_z^2\cos^2\zeta\right) \right].
\end{equation*}
This is plotted in Fig.~\ref{fig:spectral_characteristics}(e) for some values of the temperature.
At absolute zero, the spectral density is nonzero only for $\omega>0$, which corresponds to the emission of the phonons.
At finite temperature, $R_\mathrm{ph}$  is also non-vanishing for $\omega<0$, as the phonon absorption becomes possible.

Next, we consider the spectral function of the system dynamics, which is expressed in a general form in Eq.~\eqref{eq:spectral-characteristic-definintion-general-form}.
For our model, the sum runs over states $\ket*{\psi_1}$ and $\ket*{\psi_2}$ (see Eq.~\eqref{eq:stany-ortogonalne-do-psi-zero})
Correspondingly, the spectral function splits into two parts and has the form
\begin{equation}
\begin{split}
    S_\mathrm{ph} (\omega) & = s_1^\mathrm{ph} (\omega) + s_2^\mathrm{ph} (\omega) \\
    & =\frac{1}{4} \sin^2\vartheta \left|\int_{-\infty}^\infty \dd t e^{i\omega t} \sin^2\phi(t)\right|^2\\
    & + \frac{1}{4} \cos^2\frac{\vartheta}{2} \left|\int_{-\infty}^\infty \dd t e^{i\omega t} e^{i\left[\Lambda_2(t) - \Lambda_1(t)\right]} \sin\left[2\phi(t)\right]\right|^2.
\end{split}
\label{eq:spectral_characteristic}
\end{equation}
As presented here, the spectral function depends on the system evolution via the tipping angle $\phi(t)$ and the dynamical phases $\Lambda_{1,2}(t)$ and has the same formal form for both control schemes.
The total error, as defined in Eq.~\eqref{eq:error-environment}, takes form
\begin{gather*}
    \delta_\mathrm{ph} = \int_{-\infty}^\infty \dd \omega R_\mathrm{ph}(\omega) S_\mathrm{ph}(\omega).
    \label{eq:error_phonons}
\end{gather*}

\begin{figure}[tb]
    \centering
    \includegraphics[width=\linewidth]{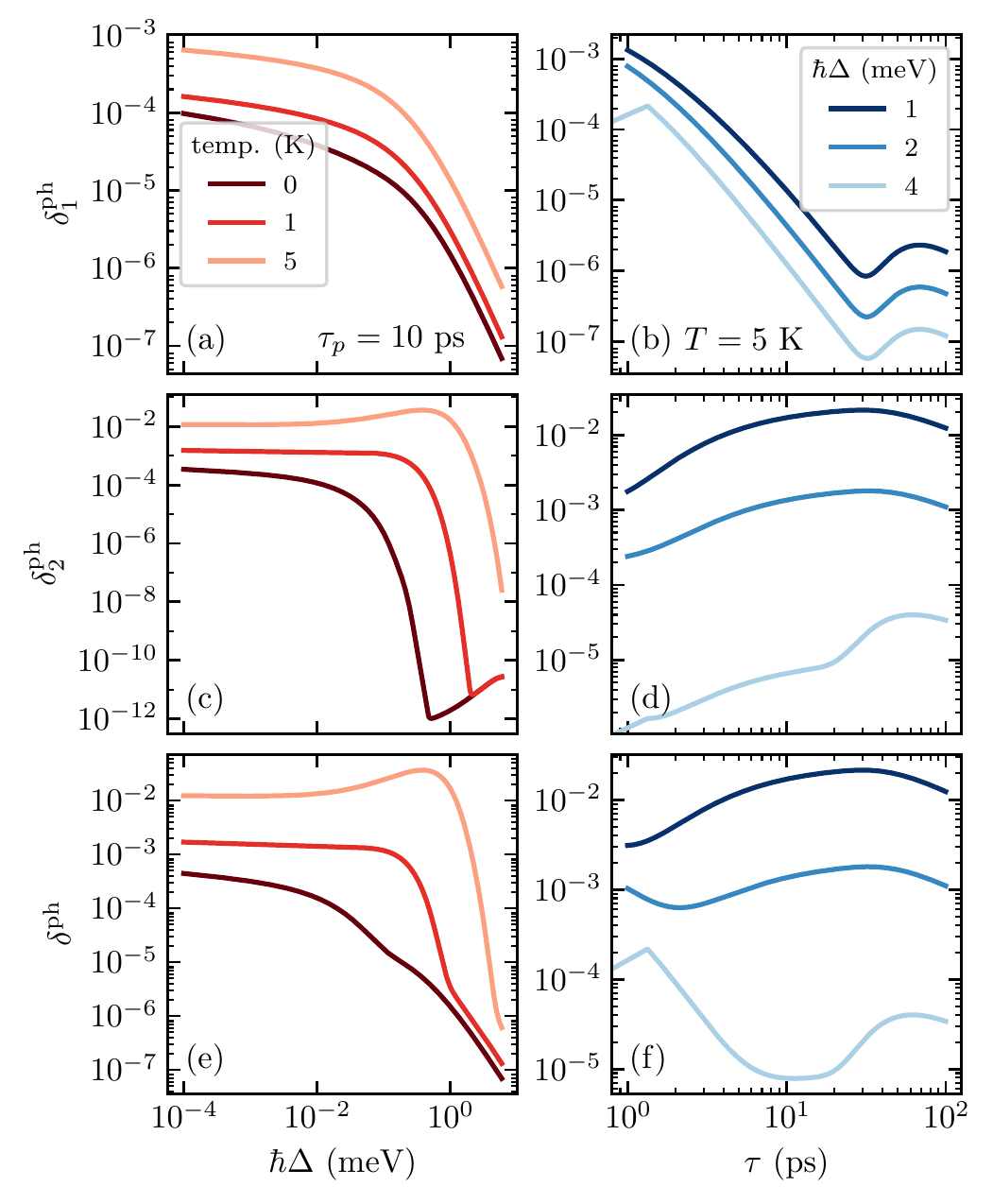}
    \caption{Gating operation error due to the influence of crystal lattice phonons  of the first kind (a,b) corresponding to the first spectral characteristic $s_1^\mathrm{ph}(\omega)$; of the second kind (c,d) corresponding to the second spectral characteristic $s_2^\mathrm{ph}(\omega)$ and the sum of them (c) as a function of the detuning $\hbar\Delta$ (a, c, e) and laser pulse duration $\tau$ (b, d, f).}
    \label{fig:error_ph}
\end{figure}

Here we investigate the value of the error averaged over all possible initial states, averaging over $\vartheta$ on the Bloch sphere and evaluating the integrals in \cref{eq:spectral_characteristic} numerically. The averaged spectral functions are denoted by $\overline{s_1^\mathrm{ph}}(\omega)$ and $\overline{s_2^\mathrm{ph}}(\omega)$.
In Fig.~\ref{fig:spectral_characteristics}(a-d), we plot these functions for selected values of the detuning and pulse duration, for the special case of $\pi$ rotation around the $x$ axis on the Bloch sphere, when the phonon response to both protocols is the same.

It is useful to derive approximate analytical expressions for the characteristic functions that are obtainable in the regime of $\Omega_0 \ll \Delta$ which occurs for sufficiently long pulses or large detunings (see Fig.~\ref{fig:tipping_angle}(a)).
The spectral functions take the approximate form (see Ref.~\cite{Grodecka2007})
\begin{subequations}
\begin{equation}
    \overline{s_1^\mathrm{ph}}(\omega) \approx \frac{\pi}{96} \frac{\Omega_0^4 \tau^2}{\Delta^4} \exp\left(-\frac{1}{2}\tau^2\omega^2\right)
    \label{eq:average_spec_char_1}
\end{equation}
and
\begin{equation}
\begin{split}
    \overline{s_2^\mathrm{ph}}(\omega) \approx & \frac{\pi}{4} \frac{\Omega_0^2\tau^2}{\Delta^2}\left\{ \exp\left[-\frac{1}{2} (\Delta + \omega)^2\right] \right.\\
    &\left. - \frac{\Omega_0^2}{2\sqrt{3}\Delta^2}\exp\left[-\frac{1}{6}\tau^2 (\Delta+\omega)^2\right]\right\}^2.
\label{eq:average_spec_char_2}
\end{split}
\end{equation}
\end{subequations}
One can understand the physical meaning of the spectral functions using Fig.~\ref{fig:spectral_characteristics} and the approximate forms given above.
First, $\overline{s_1^\mathrm{ph}}$ is a symmetric function of $\omega$, centered at $\omega=0$, and broadens with decreasing $\tau$. It corresponds to dynamically induced pure dephasing processes and overlaps considerably with the spectral density for small pulse duration even at zero temperature. However, its area diminishes as detuning increases.
Thus, the corresponding error $\delta_1$ decreases with increasing pulse duration and diminishes with the value of detuning as is clear from Fig.~\ref{fig:error_ph}(a,b).
Second, the function $\overline{s_2^\mathrm{ph}}(\omega)$ has two peaks of which the large one is centered around $\omega=-\Delta$ and broadens with decreasing pulse duration, while the small one is located to the right of the former having a width independent of both $\tau$ and $\Delta$. This spectral structure corresponds to the phonon-assisted generation of the excited state by absorbing a phonon to provide the missing energy $\Delta$ (for detuning below the optical transition). This process has an approximately resonant nature, with a broadening due to the driven dynamics of the system.

\subsection{Error due to radiative recombination \label{sec:photon-bath}}

In this subsection, we consider the error originating from the spontaneous emission from the state $\ket*{X}$ (radiative recombination), which is slightly occupied during the gating procedure and induced by the interaction Hamiltonian in Eq.~\eqref{eq:carrier-photon-interaction}.
We apply the theory presented in Sec.~\ref{sec:general-theory of-error} in a similar way to the phonon error studied in Sec.~\ref{sec:phonon-bath}.

The system and reservoir operators in Eq.~\eqref{eq:general-form-of-interaction} in the interaction picture with respect to the unperturbed evolution have the form
\begin{align*}
\begin{aligned}
\widehat{S}_{\mathrm{rad}}(t) =&U^\dag_c(t) 
\left[e^{-i\omega_T t}\right. \dyad{T}{X}\\
&\left.+ e^{-i\omega_S t}\dyad{S}{X}\right]
U_\mathrm{c}(t) + \mathrm{h.c.} 
\end{aligned}
\end{align*}
and 
\begin{equation*}
\widehat{R}_{\mathrm{rad}}(t) =  \sum_{\mathbf{q}, \lambda} g_{\mathbf{q}\lambda} c_{\mathbf{q}\lambda}^\dagger
+\mathrm{h.c.}
\end{equation*}
The radiative reservoir is assumed to be in the vacuum state, hence the corresponding spectral density is nonzero only at positive frequencies and only the positive-frequency part of the spectral functions contributes. According to Eq.~(\ref{eq:spectral-characteristic-definintion-general-form}), the spectral function can be written as 
\begin{equation*}
    S_\mathrm{rad}(\omega) = \sum_i \left|\mel{\psi_i}{Y_\mathrm{rad}(\omega)}{\psi_0}\right|^2,
    \label{eq:radiation-spectral-characteristic}
\end{equation*}
where $Y_\mathrm{rad}(\omega)$ is defined in Eq.~(\ref{eq:error-y-function}).
The system operator in the present case involves interband transitions, hence the spectral function is centered around the interband transition frequency $\omega_0$, which is on the order of fs$^{-1}$, while its dynamically-induced broadening is on the order of ps$^{-1}$ (the typical time scale of the qubit dynamics). Since the radiative spectral density is a smooth function, one can approximate the result by a Markovian formula
\begin{align}
\begin{aligned}
   \delta_\mathrm{rad} &= \int_{-\infty}^\infty \dd \omega  R_\mathrm{rad}(\omega) S_\mathrm{rad}(\omega)\\ 
& = \Gamma \frac{1}{2\pi} \int_{-\infty}^\infty \dd \omega   S_\mathrm{rad}(\omega), 
\end{aligned}
\label{err-rad}
\end{align}
where $\Gamma=2\pi R(\omega_0)$ is the spontaneous emission (radiative recombination) rate, which is known from experiment. One can write
\begin{equation*}
\mel{\psi_i}{Y_\mathrm{rad}(\omega)}{\psi_0} = 
  \int \dd t e^{i\omega t} s_i^\mathrm{rad}(t),
\end{equation*}
where 
$s_i^\mathrm{rad}(t)=\mel{\psi_i}{S_{\mathrm{rad}}(t)}{\psi_0}$. Then
\begin{equation*}
\frac{1}{2\pi}\int_{-\infty}^\infty \dd \omega   \widetilde{S}_\mathrm{rad}(\omega) = 
 \sum_i \int_{-\infty}^\infty \dd t \left| s_i(t) \right|^2.
\end{equation*}
For the two-pulse protocol, the explicit forms of the two functions involved are 
\begin{align*}
    s_1^\mathrm{rad}(t) =& - \frac{1}{4} \sin{\vartheta}\sin{\left[2\phi(t) \right]}\\
    &\times\left(e^{- i \omega_{S} t} \sin{\beta} + e^{- i \left(\omega_{T} t + \gamma\right)} \cos{\beta}\right)\\
    &- \cos^2{\frac{\vartheta}{2}} \sin{\left[\phi(t) \right]}  e^{- i \Lambda_{1}(t)} e^{- i \varphi} \\
    &\times \left(e^{- i \omega_{S} t} \cos{\beta} - e^{- i \left(\gamma + \omega_{T} t\right)} \sin{\beta}\right)
\end{align*}
and 
\begin{align*}
\begin{aligned}
    s_2^\mathrm{rad}(t) =& -\cos{\frac{\vartheta}{2}} \sin^{2}{\left[\phi(t) \right]} e^{i \left[\Lambda_{2}(t)-\Lambda_{1}(t)\right]}  \\
    &\times\left[e^{- i \omega_{S} t} \sin{\beta} + e^{- i \left(\omega_{T} t + \gamma\right)} \cos{\beta}\right].
\end{aligned}
\end{align*}
The functions for the single-pulse protocol are obtained by restricting these equations to rotations around the $x$ axis, i.e., $\beta=\pi/4$, $\gamma=0$.

 \begin{figure}[tb]
    \centering
    \includegraphics[width=\linewidth]{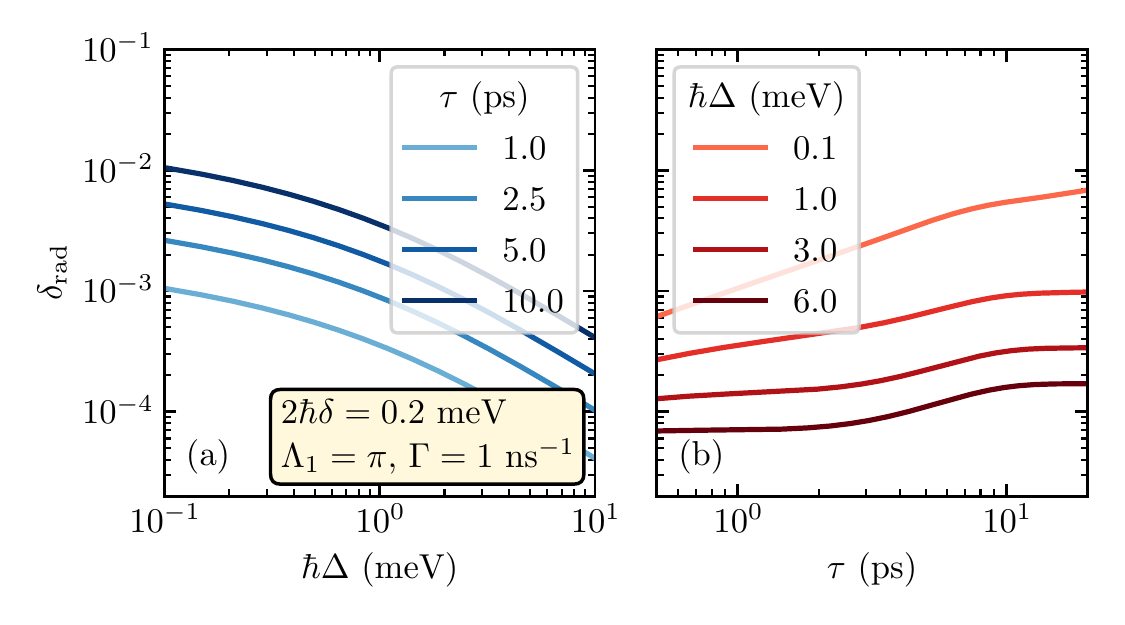}
    \caption{Radiative recombination error: (a) as a function of pulse detuning and (b) pulse duration
     for a $\pi$ rotation  around the $x$-axis, averaged over initial states. For such a choice of the rotation axis, this error is identical in both protocols. We assume  $2\hbar\delta = 0.2$~meV and $\Gamma=1$~ns$^{-1}$.}
    \label{fig:photon-error}
\end{figure}

The radiative contribution to the error, defined by Eq.~(\ref{err-rad}), for a $\pi$ rotation around the $x$ axis and averaged over initial states, is shown in Fig.~\ref{fig:photon-error} for the exchange splitting of $0.2$~meV and an exciton lifetime of 1~ns.
The error is reduced by increasing pulse detuning, which results in a weaker occupation of the $\ket*{X}$ state. Since this occupation is small at large detunings (see Fig.~\ref{fig:na+off-resonant}), the error is on the order of $10^{-4}$ even though the duration of the gating reaches 10\% of the exciton lifetime. Although in the Markov approximation the probability of spontaneous emission accumulates with time, the resulting error shows a sublinear dependence on the pulse duration at sufficiently large detunings, since for a longer pulse duration the pulse amplitude $\Omega_0$ is lower for a given rotation angle, which reduces the occupation of the excited state during the evolution. 

\section{Summary and conclusions \label{sec:summary-and-discussion}}

We have presented an extensive analysis of the accuracy and dephasing effects on two possible optical Raman control schemes for a singlet-triplet qubit encoded in a two-electron system in a self-assembled QD. 

Our study reveals the dependence of the fidelity of quantum control on the system and driving parameters: the exchange splitting in the two-electron system, the detuning from the optical transition, and the pulse duration. The results show that high-fidelity operation requires appropriate optimization of these parameters, depending on the control scheme.
In both protocols, the probability of leakage to the auxiliary excited state becomes small for pulse durations of a few picoseconds or longer and decreases with the growing detuning. Phonon induced errors are small for pulse durations of at least a few picoseconds or as long as the detuning of the optical coupling is sufficiently large.
For a protocol based on two spectrally selective optical pulses, the rotation accuracy within the qubit subspace is very small (10$^{-3}$ or lower) for pulse durations of several picoseconds, which is consistent with reducing the leakage but incurs a higher impact from spontaneous emission. The latter, however, can be suppressed to values on the order of 10$^{-4}$ by increasing the detuning of the optical coupling. For a high-accuracy single-pulse scheme, which assumes nonselective optical coupling, short pulses are preferred for lowering the gating error, which creates a trade-off situation against the leakage error. This trade-off can be mitigated by the selection of optical detunings and using systems with small exchange splitting.

In general, large optical detuning (which is an easily tunable parameter of the experiment) is favorable for both protocols, while the two schemes require the opposite engineering of the exchange splitting (which depends on the structure morphology and can be tuned to some extent with an external electric field). 

\acknowledgments
We acknowledge support from the Polish National Science Centre (NCN) under Grant no. 2016/23/G/ST3/04324 (K.K. and P.M.) and NAWA APM Grant no. PPI/APM/2019/1/00085/U/00001 (K.K.).


%
\end{document}